\documentclass[%
 reprint,
superscriptaddress,
nofootinbib,
 amsmath,amssymb,
 aps,
floatfix,
]{revtex4-2}

\usepackage{graphicx}
\usepackage{dcolumn}
\usepackage{bm}
\usepackage{algorithm}
\usepackage{algpseudocode}
\usepackage{amsmath}
\usepackage{braket}
\usepackage[all,pdf]{xy}
\usepackage[qm]{qcircuit}
\usepackage{physics}
\usepackage[colorlinks=true,citecolor=blue,linkcolor=magenta]{hyperref}
\usepackage{multirow}
\usepackage{appendix}
\usepackage{booktabs}

\makeatletter
\def\caption@documentclass{elsarticle} 
\makeatother

\usepackage[caption=false]{subfig}

\newcommand{\nc}{\newcommand}
\nc{\disp}{\displaystyle}
\nc{\argmax}{\mathop{\rm arg~max}\limits}
\nc{\argmin}{\mathop{\rm arg~min}\limits}
\nc{\lr}[1]{\ensuremath{\left(#1\right)}}
\nc{\mlr}[1]{\mleft(#1\mright)}
\nc{\alr}[1]{\ensuremath{\left\langle#1\right\rangle}} 
\nc{\blr}[1]{\ensuremath{\left[#1\right]}} 
\nc{\clr}[1]{\ensuremath{\left\{#1\right\}}} 
\nc{\nlr}[1]{\ensuremath{\left\|#1\right\|}}
\nc{\vlr}[1]{\ensuremath{\left|#1\right|}}
\nc{\amlr}[1]{\mleft\langle#1\mright\rangle} 
\nc{\bmlr}[1]{\mleft[#1\mright]} 
\nc{\cmlr}[1]{\mleft\{#1\mright\}} 
\nc{\nmlr}[1]{\mleft\|#1\mright\|}
\nc{\vmlr}[1]{\mleft|#1\mright|}

\nc{\rmvec}{\mathop{\rm vec}}
\nc{\rmtr}{\mathop{\rm tr}}
\nc{\tp}{\mathsf{T}}
\nc{\ex}{\mathbb{E}}
\nc{\defeq}{:=}
\nc{\cpad}{~}
\nc{\setdef}[2]{\clr{#1 \mathrel{}\middle|\mathrel{} #2}}
\nc{\Z}{\mathbb{Z}}
\nc{\R}{\mathbb{R}}
\nc{\N}{\mathbb{N}}
\nc{\ra}{\rightarrow}

\algnewcommand{\Initialize}[1]{%
  \State \textbf{Initialize: }
  \State \hspace*{\algorithmicindent}\parbox[t]{0.8\linewidth}{\raggedright #1}
}

\begin{document}

\preprint{APS/123-QED}

\title{Continuous optimization by quantum adaptive distribution search}

\author{Kohei Morimoto}
 \email{morimoto.kohei.p55@kyoto-u.jp}
\affiliation{   
Graduate School of Informatics, Kyoto University, Yoshida-honmachi, Sakyo-ku, Kyoto 606-8501, Japan}
\author{Yusuke Takase}%
 \email{y.takase@sys.i.kyoto-u.ac.jp}
\affiliation{%
    Graduate School of Informatics, Kyoto University,
    Yoshida-honmachi, Sakyo-ku, Kyoto 606-8501, Japan
}%
\author{Kosuke Mitarai}%
 \email{mitarai.kosuke.es@osaka-u.ac.jp}
\affiliation{%
  Graduate School of Engineering Science, Osaka University, 1-3 Machikaneyama, Toyonaka, Osaka 560-8531, Japan
}%
\affiliation{%
  Center for Quantum Information and Quantum Biology,
  Osaka University, 1-2 Machikaneyama, Toyonaka 560-0043, Japan
}
\author{Keisuke Fujii}
 \email{fujii@qc.ee.es.osaka-u.ac.jp}
\affiliation{%
  Graduate School of Engineering Science, Osaka University, 1-3 Machikaneyama, Toyonaka, Osaka 560-8531, Japan
}%
\affiliation{%
  Center for Quantum Information and Quantum Biology,
  Osaka University, 1-2 Machikaneyama, Toyonaka 560-0043, Japan
}%
\affiliation{%
  RIKEN Center for Quantum Computing (RQC),
  Hirosawa 2-1, Wako, Saitama 351-0198, Japan
}%

\date{\today}

\begin{abstract}
In this paper, we introduce the quantum adaptive distribution search (QuADS), a quantum continuous optimization algorithm that integrates Grover adaptive search (GAS) with the covariance matrix adaptation - evolution strategy (CMA-ES), a classical technique for continuous optimization.
QuADS utilizes the quantum-based search capabilities of GAS and enhances them with the principles of CMA-ES for more efficient optimization.
It employs a multivariate normal distribution for the initial state of the quantum search and repeatedly updates it throughout the optimization process. 
Our numerical simulations show that QuADS outperforms both GAS and CMA-ES. 
This is achieved through adaptive refinement of the initial state distribution rather than consistently using a uniform state, resulting in fewer oracle calls.
This study presents an important step toward exploiting the potential of quantum computing for continuous optimization.
\end{abstract}

\maketitle


\section{INTRODUCTION}
Continuous optimization has become a critical element across a wide range of fields in recent times \cite{nocedal1999numerical, griva2008linear}.
However, the challenge of global optimization of non-convex functions remains an unresolved issue that demands further exploration \cite{hendrix2010introduction, horst2000introduction, zhigljavsky2007stochastic}. 
The potential benefits of conquering this challenge are immense, and efficient optimization methods for non-convex functions lead to breakthroughs in numerous fields. 

In quantum computing, researchers have been investigating the potential of quantum computers in addressing continuous optimization problems. 
Grover Adaptive Search (GAS) \cite{durr_quantum_1999, baritompa_grovers_2005}, designed for fault-tolerant quantum computers (FTQC), has emerged as a notable optimizer for discrete problems. 
This algorithm utilizes Grover's algorithm \cite{grover_fast_1996, boyer_tight_1998-1} to mark regions where the function value is below the current optimum and iteratively finds superior solutions.
Furthermore, the extension of GAS for the continuous optimization problem has been proposed in Refs. \cite{protopopescu_solving_2002, protopopescu_quantum_2005}. 
Since Grover's algorithm assures a quadratic improvement in performance with respect to the number of oracle calls, every update step of the GAS can also achieve such a speedup.
 
However, the vanilla GAS is not the best choice for the optimization problems commonly found in practical situations, as it employs a uniform search across the entire domain of the function.
Bulger \cite{bulger_quantum_2005, bulger_combining_2007} has proposed integrating local search within the oracle to concentrate the search points around the local optimum within each basin.
While it achieves considerable speedup for some cases with respect to the oracle calls, the coherent arithmetics to perform the local search generally increase the cost to implement the oracle itself, and thus, the overall speedup over the vanilla GAS is not apparent. Additionally, this algorithm does not take into account the structure of the function outside of individual basins, although some optimization problems could benefit from considering their overall structure.

In this article, we propose an extension of GAS without modifying the oracle but by modifying the initial state, thus avoiding making the oracle complex in contrast to Refs. \cite{bulger_combining_2007, bulger_quantum_2005}.
Namely, we adaptively update the initial state of GAS during the optimization process, utilizing the information about the objective function obtained from the optimization history. 
In the updating rule, we incorporate the concept of a classical optimization method known as covariance matrix adaptation - evolution strategy (CMA-ES) \cite{hansen_cma_2023}. CMA-ES is known for efficiently minimizing unimodal test functions and is also highly competitive in the global optimization of multimodal functions \cite{hansen2004ecm}, making it widely applicable in various fields, including generative models \cite{ha_recurrent_2018} and tensor network models \cite{rudolph2023synergy}. 

Inspired by the idea of adapting probability distributions in CMA-ES, we propose a natural extension of GAS, which we call quantum adaptive distribution search (QuADS), that combines Grover adaptive search and CMA-ES principles. 
Our algorithm aims to optimize continuous functions by adaptively changing the belief distribution and using the quantum-based search capabilities of amplitude amplification. 
Through numerical simulations in scenarios of up to 10 dimensions, we observe that GAS constantly needs high computational costs regardless of the type of function instead of the assurance of finding the global optimum. 
Also, CMA-ES typically converges faster than other methods, but it frequently falls into local optima. 
QuADS consistently outperforms GAS and performs better than CMA-ES, especially for high dimensional cases in terms of expected oracle counts until obtaining the global optimal solution. 
We believe QuADS will be one of the promising tools for accelerating continuous optimization on quantum computers.

The rest of this paper is organized as follows.
In Section \ref{sec:preliminaries}, we provide summaries of GAS and CMA-ES.
In Section \ref{sec:quads}, we detail the main concepts and algorithms of QuADS. In Section \ref{sec:experiments}, we validate our method through numerical simulations and show that the proposed method surpasses traditional approaches.
Finally, we conclude the paper in Section \ref{sec:conclusion}. 
Code for our numerical simulation can be found at Ref. \cite{quads_github}.



\section{Preliminaries}
\label{sec:preliminaries}

\subsection{Grover Adaptive Search}


Grover's algorithm~\cite{grover_fast_1996, boyer_tight_1998-1} achieves a quadratic speedup in searching for specific items within unstructured tables.
Specifically, given a set of elements $A = \{a_i\}$ $(i=0, \dots, n-1)$ and a subset of 'good' elements $G\subset A$, Grover's algorithm efficiently identifies the index of the element that belong to $G$. 
Grover's algorithm is a special case of the more general concept of amplitude amplification~\cite{brassard_quantum_2002}, where the initial state is uniform. 
The algorithm for amplitude amplification is shown in Algorithm~\ref{alg:amplitude-amplification}.
$O$ represents an oracle gate, which selectively inverts the phase of the amplitude of states corresponding to the 'good' elements.
$\lambda$ is selected in the range $(1, 4/3)$ because it is guaranteed to achieve quadratic speedup within this range~\cite{boyer_tight_1998-1}.

Grover Adaptive Search~(GAS)~\cite{durr_quantum_1999, baritompa_grovers_2005} is a quantum optimization method that utilizes Grover's algorithm.
GAS begins with a uniform superposition state and, using the current minimum value as a threshold, amplifies the amplitudes of states below this threshold through Grover's search.
See Algorithm~\ref{alg:grover} for the pseudo-code of the algorithm. $O_{f,\theta}$ is an oracle with objective function $f$ to perform a phase flip in regions where $f$ is less than the threshold $\theta$ and $H$ denotes Hadamard gate.

GAS has been extended to continuous optimization problems in Refs.~\cite{protopopescu_solving_2002, protopopescu_quantum_2005}. 
It is done by treating continuous optimization problems the same way as discrete ones by discretizing the continuous space using fixed-point representation. 
In continuous problems, the number of discretization grids is $2^{D\tau}$, where $D$ is the function's dimension, and $\tau$ is the bit count for the continuous variable.
This technique is especially effective for optimization problems of highly discontinuous functions due to the inherent nature of Grover's algorithm, which accelerates the search in ``needle in a haystack'' scenarios.
However, in practical situations, GAS is not the best choice because employing methods beyond uniform search is advantageous.

\begin{figure}[!t]
\begin{algorithm}[H]
\caption{Amplitude amplification}
\label{alg:amplitude-amplification}
\begin{algorithmic}
    \Require state preparation operator $P_0$, oracle $O$
    \Initialize{
        Set $m=1$, parameter $\lambda \in (1, 4/3)$.
    }
    \Repeat
        \State Choose a rotation count $r$ uniformly from $\{0, \dots, \lfloor m \rfloor \}.$
        \State Prepare initial state: $\ket{\psi} = P_0 \ket{0}$
        \State Apply amplitude amplification with $r$ rotation: 
        \State \qquad $\ket{\psi} = \left(P_0 \left(I-2\ket{0}\bra{0}\right) P_0^\dag O\right)^r \ket{\psi}$
        \State Observe $\ket{\psi}$ and let $x$ be outcome.
        \State Increase rotation count: $m = m \lambda.$
    \Until{good element is found}
\end{algorithmic}
\end{algorithm}

\begin{algorithm}[H]
\caption{Grover Adaptive Search}
\label{alg:grover}
\begin{algorithmic}
    \Require Initial point $\mu_0$, oracle $O_{f,\theta}$
    \Initialize{
        Set the initial threshold $\theta \leftarrow f(\mu_0)$ for the initial point.
    }
    \Repeat
        \State $x \leftarrow \mathrm{AmplitudeAmplification}(H^{\otimes D}, O_{f,\theta})$
        \State $\theta \leftarrow f(x)$
    \Until{the termination criterion is met}
\end{algorithmic}
\end{algorithm}

\begin{algorithm}[H]

\caption{CMA-ES}
\label{alg:CMAES}
\begin{algorithmic}
%
    \Require $\mu_0$, $C_0$, $\sigma_0$, number of samples $M$, number of best individuals $K$


    \Repeat
        \State Sample $M$ independent samples $x_i ~ \lr{i=1, \ldots, M}$ from $\mathcal{N}(\mu, \sigma^2 C).$
        \State Choose the best $K$ individuals $X=\clr{X_j}_{j=1}^{K}$ from $\clr{x_i}_{i=1}^{M}.$
        \State $(\mu, C, \sigma, p_c, p_\sigma) \leftarrow \mathrm{CMA\mathchar`-ES~ update}(\mu, C, \sigma, p_c, p_\sigma; X)$
    \Until{convergence}
\end{algorithmic}
\end{algorithm}

\end{figure}

\subsection{CMA-ES}

\begin{figure}[thb]
\begin{algorithm}[H]
\caption{CMA-ES update \cite{hansen_cma_2023}}
\label{alg:cma-update}
\begin{algorithmic}
    \Require selected samples $X \in \mathbb{R}^{K \times D}$ (X is rearranged like $f(X_i) \leq f(X_j)$ for $i \leq j $), mean $\mu$, scaled covariance $C$, step size $\sigma$, evolution paths $p_c$, $p_\theta$, hyper parameters ($c_\sigma$, $c_c$, $\mu_w$, $w_{i=1...\mu}$)
    \Ensure $\mu$, $C$, $\sigma$, $p_c$, $p_\theta$
    \\
    \State Normalize samples using previous mean and step size:
    \begin{align*}
        Z_i = \frac{X_i - \mu}{\sigma}.
    \end{align*}
        
    \State Calculate the weighted mean of $Z_i$, assigning larger weights to superior samples:
    \begin{align*}
    \ev{Z}_{w} = \sum_{i=1}^K w_i Z_i.
    \end{align*}
    
    \State Update $\mu$ by weighted mean of $X$:
    \begin{align*}
        \mu \leftarrow \sum_{i=1}^K w_i X_i.
    \end{align*}

    \State Update $p_\sigma$ to track how much the samples have moved from the mean ($\mu_{\mathrm{eff}} = 1/\sum_i w_i^2$):
    \begin{align*}
    p_\sigma \leftarrow (1-c_\sigma) p_\sigma + \sqrt{c_\sigma(2-c_\sigma)\mu_{\mathrm{eff}}} C^{-1/2} \ev{Z}_{w}.
    \end{align*}
    \State Adjust the step size to keep the length of the evolved path close to $\mathbb{E}\norm{\mathcal{N}(\bm 0, \bm I)}$:
    \begin{align*}
        \sigma \leftarrow \sigma \exp\left(\frac{c_\sigma}{d_\sigma} \left(\frac{\norm{ p_\sigma}}{\mathbb{E}\norm{\mathcal{N}(\bm 0, \bm I)}}-1\right)\right).
    \end{align*}

    \State Define $h_\sigma^g$ where $g$ denote current iteration number. $h_\sigma^g$ is used to prevent $C$ from increasing too fast.
    \begin{align*}
        h_\sigma := 
    \begin{cases} 
    1 & \text{if } \frac{\norm{p_\sigma}}{\sqrt{1-(1-c_\sigma)^{2(g+1)}}}
 < (1.4 + \frac{2}{n+1})\mathbb{E} \norm{N(0, 1)} \\
    0 & \text{otherwise}
    \end{cases}
    \end{align*}
     
    \State Update $p_c$ to keep the average direction of $Z$.
    \begin{align*}
        p_c \leftarrow (1-c_c) p_c + h_\sigma \sqrt{c_c(2-c_c)\mu_{\mathrm{eff}}} \ev{Z}_w.
    \end{align*}

    \State Update $C$ by average direction and one-step covariance:
    \begin{align*}
    \begin{split}
        C \leftarrow  &\lr{1 + c_1 \delta(h_\sigma) -  c_1 - c_\mu\sum_{j=1}^K w_j} C \\ &+ c_1 p_c p_c^T + c_\mu \sum_{i=1}^K w_i Z_i Z_i^T.
    \end{split}
    \end{align*}

\end{algorithmic}
\end{algorithm}
\end{figure}

CMA-ES \cite{hansen_cma_2023} is a black-box optimization algorithm based on the evolution strategy \cite{back1994basic} for complex and non-convex functions. 
Compared to optimization methods using derivatives, this method is more robust to multimodal functions \cite{hansen2004ecm}.
Also, the well-designed parameter update rule makes the algorithm invariant to coordinate affine transformations and be able to handle bad scaling where the scale of the function varies for each dimension.

We briefly describe the CMA-ES algorithm (Algorithm~\ref{alg:CMAES}). This method uses a multivariate normal distribution for the search process, denoted by $\mathcal{N}(\mu, \sigma^2 C)$, where $\mu$ is the mean, $C$ is the scaled covariance, and $\sigma$ is the step size. While $C$ and $\sigma$ contribute to defining the covariance, $C$ primarily determines the shape and orientation, whereas $\sigma$ determines the scale of the search. The algorithm proceeds through the following iterative steps:
Initially, $M$ search points, $x_1, \ldots, x_M$, are generated from the normal distribution $\mathcal{N}(\mu, \sigma^2 C)$. Subsequently, the function values $f(x_1), \ldots, f(x_M)$ at each candidate point are evaluated, and the top $K (< M)$ individuals with the lowest function values are selected.
Then, the parameters of the normal distribution are updated accordingly.
$\mu$ is adjusted toward the average of the best individuals. $C$ is updated to reflect the direction in which the function value changes significantly, and $\sigma$ increases when $\mu$ consistently moves in a similar direction over several iterations and decreases when the movement of $\mu$ appears more random. 
For updating $C$ and $\sigma$, the evolution path (moving average) $p_c$ and $p_\sigma$ are utilized.
The detailed parameter update rule is shown in Algorithm \ref{alg:cma-update} (see Ref.~\cite{hansen_cma_2023} for further detail). 
By repeating these operations, the normal distribution progressively approaches the optimal solution.

\section{Quantum adaptive distribution search}
\label{sec:quads}

\begin{figure}[thbp]
\begin{algorithm}[H]
\caption{QuADS (proposed method)}
\label{alg:proposed}
\begin{algorithmic}
%
\Require initial mean $\mu_0$, initial covariance $C_0$, initial step size $\sigma_0$, initial threshold $\theta$, oracle $O_{f,\theta}$, number of samples $K$, smoothing factor $\alpha$, quantile $q$

    \Repeat
        \State $X, Y \leftarrow \emptyset$
        \Repeat
            \State $x \leftarrow \mathrm{Amplitude ~amplification}(\mathcal{G}_{\mu,\sigma^2 C}, O_{f,\theta}, \theta)$
            \State $X \leftarrow X \cup \{x\}$
            \State $Y \leftarrow Y \cup \{f(x)\}$
        \Until {$\vlr{X} = K$}
        \State $(\mu, C, \sigma, p_c, p_\sigma) \leftarrow \mathrm{CMA\mathchar`-ES~update}(\mu, C, \sigma, p_c, p_\sigma; X)$
        \State $\theta \leftarrow \alpha \theta + (1-\alpha) \mathrm{quantile}_{q}(Y)$
    \Until{convergence}
\end{algorithmic}
\end{algorithm}
\end{figure}

The main concept of our method is to extend GAS by using an adaptive multivariate normal distribution for the initial state in quantum search.
Denote $p$ is the probability of finding a point below a specified threshold in the distribution.
Compared to GAS approach, QuADS is expected to achieve a higher $p$ by initiating the quantum search with an adaptive distribution instead of a uniform distribution. 
This is especially advantageous when the threshold is near the global optimum because while GAS consistently demands computational resources proportional to $O(\sqrt{2^{D\tau}})$, QuADS has the potential to reduce this cost when the adaptive distribution is properly updated toward the optima.
From another perspective, our method enhances the efficiency of the sampling process in the CMA-ES by using amplitude amplification to sample from regions where function values are below an adaptive threshold.
Classically, the number of oracle calls required to find a sample is $O(1/p)$. 
However, amplitude amplification can reduce this to $O(1/\sqrt{p})$ oracle calls. 
It should be noted that QuADS may occasionally converge on local optima due to its adaptive distribution updating strategy, in contrast to GAS, which always identifies the global optimum.
Nonetheless, in terms of the expected oracle call to locate the global optimum, we find QuADS outperforms GAS, as will be discussed in detail in Section \ref{sec:experiments}. 

\begin{figure}[tbp]
    \input{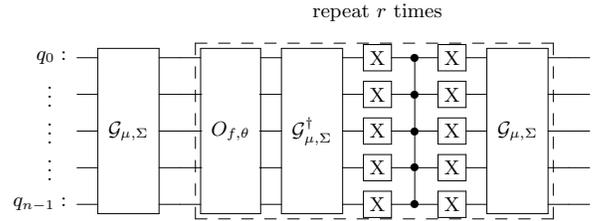}
    \caption{Structure of the quantum circuit for the search part of QuADS. $\mathcal{G}_{\mu, \Sigma}$ gate represents the circuit for preparing a normal distribution. The dashed block is repeated $r$ times, where $r$ represents the rotation number for amplitude amplification.
    In QuADS, we iteratively perform sampling from this circuit using random $r$ and subsequently update the parameters $\mu$ and $\Sigma$ by the samples.
    }
    \centering
    \label{fig:circuit-quads}
\end{figure}

Algorithm \ref{alg:proposed} describes the QuADS, where $\mathcal{G}_{\mu, \Sigma}$ is a preparation operator for a multivariate normal distribution with mean $\mu$ and covariance $\Sigma$, that is,
\begin{align*}
    \mathcal{G}_{\mu, \Sigma} \ket{0^{D\tau}} = Z \sum_{x} \sqrt{ e^{-\frac{1}{2}(x-\mu)^{\mathrm{T}} \Sigma^{-1}(x-\mu)}} \ket{x},
\end{align*}
where $Z$ is the normalization constant.
The construction of $\mathcal{G}_{\mu, \Sigma}$ has been discussed in Refs. \cite{kitaev_wavefunction_2009, bauer_practical_2021}. This operator can be implemented using a polynomial number of quantum gates with respect to $D$ and $\tau$.
Our approach begins by sampling $K$ superior points from the region below the adaptive threshold $\theta$. In the sampling, amplitude amplification is used, with the normal distribution serving as an initial state.
Figure~\ref{fig:circuit-quads} illustrates the corresponding quantum circuit for the amplitude amplification.
These sampled points are then used to update both the thresholds and the normal distribution parameters.
For updating the distribution parameters, we employ the same updating rule as in CMA-ES.
By utilizing the updating rule from CMA-ES, we can combine insights about the structure of the function and lead to an efficient search.
As for the updating rule for thresholds, QuADS uses a moving average of the $q$-quantile of the function value at the sampled points, i.e., approximately equivalent to considering the best $Mq$th value out of $M$ samples, for some constant $q$.
The reason for using this threshold update rule rather than using the minimum of sampled values as done in GAS is that if the threshold is updated prematurely, the distribution may not have sufficient weights on the region below the threshold, and that causes inefficient sampling. 

\begin{figure}[tbp]
\includegraphics[scale=0.33]{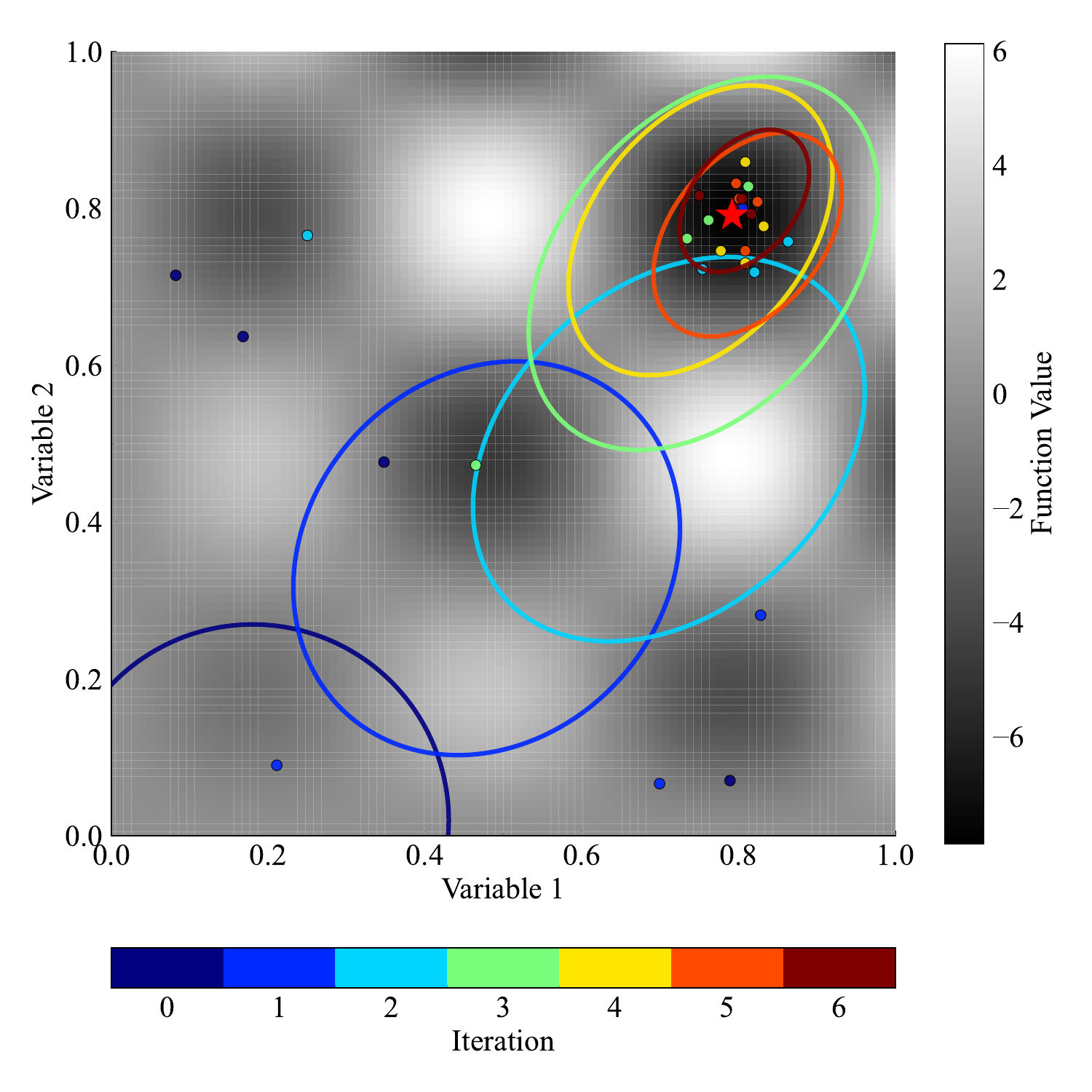} 
\caption{\label{fig:algorithm_alpine} Typical optimization process of QuADS. 
The solid line represents the $1\sigma$ region of the distribution, and the points represent the samples from amplitude amplification at each iteration. The star mark represents the global optimum point.
}
\end{figure}

Fig.~\ref{fig:algorithm_alpine} shows an example of the QuADS optimization process for the alpine02 function in Table \ref{tab:test-func}. 
The mean of the normal distribution progressively shifts toward the global optimum, while the distribution's covariance broadens in the direction of this movement.
Each iteration's points are sampled through amplitude amplification using the normal distribution as a prior.
Notably, more samples are drawn from areas where the normal distribution's weights are larger, within regions with small function values.

\section{Numerical Simulations}
\label{sec:experiments}

\subsection{Performance estimation by amplitude simulation}

In this section, we compare the performance of QuADS, GAS, and CMA-ES. 
QuADS and GAS are simulated on a classical computer.
The direct circuit simulation of these algorithms would need many ancilla qubits for performing $\mathcal{G}_{\mu, \Sigma}$ and $O_{f,y}$.
To avoid such a situation, we simulate the effect of each operation directly.
More precisely, the simulation process is conducted as follows:
We first initialize the wave function as
\begin{align*}
    \ket{\psi_0}=Z\sum_{x} \sqrt{ \exp \lr{-\frac{1}{2}(x-\mu)^{\mathrm{T}} \Sigma^{-1}(x-\mu)}} \ket{x}.
\end{align*}
Let the Grover iteration be $W = P_0 \left(I-2\ket{0}\bra{0}\right) P_0^\dag O_{f,y}$ and let $\ket{\psi_n}=W^n\ket{\psi_0}$.
If we write $\ket{\psi_{n-1}}$ as
\begin{align*}
    \ket{\psi_{n-1}}=\sum_{x=1}^{2^{D\tau}}{p_{n-1}\left(x\right)\ket{x}},
\end{align*}
then $\ket{\psi_n}$ can be calculated through two steps.
First, flip the signs of the points whose function values are below the threshold $\theta$:
\begin{align*}
    \sum_{x=1}^{2^{D\tau}}{\mathrm{sgn}\left(f(x)-\theta\right)p_{n-1}\left(x\right)\ket{x}}.
\end{align*}
Subsequently, perform a reflection on the initial state:
\begin{align*}
    \ket{\psi_n}=2\braket{\psi_{n-1}}{\psi_0} \ket{\psi_0}-\ket{\psi_{n-1}}.
\end{align*}
Using this strategy, we can simulate up to $D=3$ and $\tau=8$ on a laptop.

For multiple test functions shown in Table \ref{tab:test-func}, we run each method for $100$ times.
We chose these test functions for their multimodal characteristics and their ability to be defined in arbitrary dimensions.
We scaled the domain of the $D$ dimensional test function to be $[0, 1]^D$.
The plots of the test functions are shown in Appendix \ref{sec:test-func}.
In CMA-ES and QuADS, we uniformly sampled the initial mean $\mu_0$ from the function domain for each trial. We initialized the covariance matrix, $C_0$, as the identity matrix, and set the initial step size, $\sigma_0$, to 0.5.
We used $f(\mu_0)$ as the initial threshold of QuADS. 
For the threshold of GAS, we used $f(x)$ as the initial threshold, where $x$ is a sample from a uniform distribution on the domain. 
The termination conditions for CMA-ES and QuADS are local convergence or finding the global optimum.  Finding the global optimum is defined as sampling a point in the $\epsilon=0.01$ neighborhood of the global optimum.
Note that the global optimum of the test functions is known.
Also, we regard the local convergence as achieved when $\sigma$ becomes smaller than $\epsilon_\sigma=0.01$. 
For updating normal distribution in QuADS and CMA-ES, we use the hyperparameters shown in Appendix \ref{sec:hyper}. 
For the threshold update in QuADS, we heuristically use $\alpha = 0.5$ and $q = 0.2$. 
For the number of samples $M$ and the number of selected samples $K$ in CMA-ES, we use the default setting in Ref. \cite{hansen_cma_2023},
\begin{align*}
    M &= 4 + 3 \lfloor \ln D\rfloor, \\
    K &= \lfloor M / 2 \rfloor.
\end{align*}
For QuADS, we use the same $K$ as CMA-ES. Finally, for amplitude amplification, we set the increase rate $\lambda$ to $6/5$, suggested in \cite{boyer_tight_1998-1}.

\begin{table*}
\centering
\caption{Function definitions and domains \label{tab:test-func}
}
\begin{tabular}{lll}
\hline\hline
Name & Function form & Domain\\
\hline
rastrigin & $\sum_{i = 1}^D \lr{x_i^2 - 10\cos\lr{2 \pi x_i}}$ & [-5.12, 5.12] \\
ackley & $ -20 \exp\lr{-0.2\sqrt{\frac1D\sum_{i = 1}^D x_i^2}} - \exp \lr{\frac1D\sum_{i = 1}^D \cos\lr{2 \pi x}}$ & [-4, 4]\\
styblinski tang  & $\frac12\sum_{i = 1}^D\lr{x_i^4 - 16x_i^2 + 5x_i}$ & [-5, 5]\\
schwefel & $\sum_{i = 1}^D\lr{x_i\sin\sqrt{\abs{x_i}}}$ & [-500, 500]\\
griewank & $\frac{1}{4000} \sum_{i=1}^{n} x_i^2 - \prod_{i=1}^{n} \cos\left(\sqrt{x_i}\right)$ & [-512, 512] \\

alpine01 & $ \sum_{i = 1}^D \abs{x_i\sin\lr{x_i}+\frac{1}{10}x_i}$ & [-10, 10]\\
alpine02 & $ -\prod_{i = 1}^D\lr{\sqrt{x_i}\sin\lr{x_i}}$ & [0, 10]\\

deflected corrugated spring & $ \frac{1}{10} \sum_{i = 1}^D\lr{x_i^2-\cos\lr{5\sqrt{\sum_{j = 1}^Dx_j^2}}}$ & [0, 10]\\

wavy & $ - \sum_{i = 1}^D\cos\lr{10x_i}\exp\lr{-x_i^2/2}$ & [-$\pi$, $\pi$]\\


\hline\hline
\end{tabular}
\end{table*}

First, we compare the methods by the expected oracle calls until global convergence is reached.
The expected numbers of oracle calls in a single run of the algorithms can be written as
\begin{align}
    o_{\rm single} = o_{\rm local} (1-p_{\rm global}) + o_{\rm global}p_{\rm global},
\end{align}
where $o_{\rm local}$ and $o_{\rm global}$ are the average number of oracle calls used to reach local or global convergences, respectively, and $p_{\rm global}$ is the probability to find the global optimum. 
Note that we can approximate it as $p_{\rm global}= n_{\rm global} / 100$ where $n_{\rm global}$ is the number of simulations in which we found the global optimum. 
Then, the expected oracle calls until global convergence can be calculated as
\begin{align}
    \label{eq:expected_total_calls}
    o_{\rm total} = o_{\rm single}/p_{\rm global},
\end{align}
since we need to run the algorithms $1/p_{\rm global}$ times on average to achieve the global optimum.

\begin{figure*}[thb]
\centering
\includegraphics[scale=0.4]{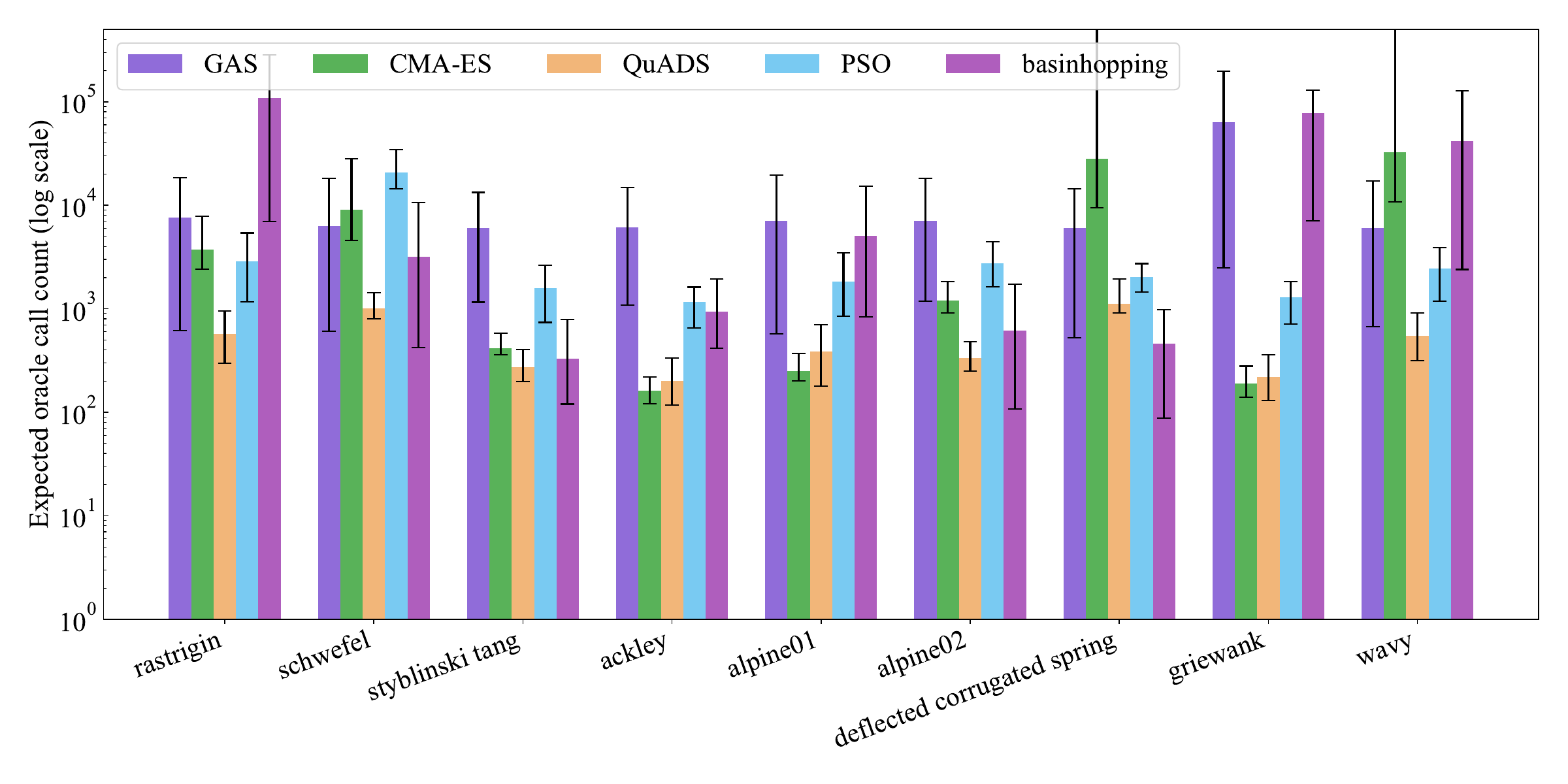} 
\caption{\label{fig:expected_global_eval}Expected oracle call counts for finding global optimum for each method (Eq. \eqref{eq:expected_total_calls}) when $D = 3$. 
We utilized the bootstrap re-sampling method to generate the error bars in these figures. 
This method involves randomly selecting data from the pool of 100 simulations and recalculating Eq. \eqref{eq:expected_total_calls} for each of these selected data sets. 
We present the estimated 5th and 95th percentiles of the expected oracle evaluation count as confidence intervals. Confidence intervals outside the figure indicate that upper bounds cannot be estimated from bootstrap. 
}

\end{figure*}

Fig. \ref{fig:expected_global_eval} shows $o_{\rm total}$ of each method.
Additionally, the figure includes results for two well-established classical optimizers: particle swarm optimization (PSO) \cite{Kennedy_particle, Shi_modified} and basinhopping optimizer \cite{wales_global_1997} for reference. The hyperparameters and implementations for the PSO and basinhopping optimizer are detailed in Appendix \ref{sec:PSO}. 
The data clearly show that QuADS consistently outperforms GAS in all cases.
For all functions except the griewank function, GAS shows almost the same number of oracle call counts. 
The reason for the increased oracle call count in the griewank function is due to its significant fluctuations around the global optimum (See Appendix \ref{sec:test-func}), resulting in a smaller success region, where the distance from the global optimum is within $\epsilon_\sigma$ and the function value is below the threshold.
Compared to classical methods CMA-ES, PSO and basinhopping, our approach demonstrates at least comparable performance and, notably, it surpasses them on rastrigin, schwefel, styblinski tang, alpine02, wavy function.
These functions, other than the rastrigin function, are characterized by multiple big valleys within the search space, each containing promising local solutions, whose optimization is difficult for CMA-ES.
On the other hand, in the case of the ackley, alpine01, and griewank functions, characterized by numerous small local optima within a large valley, QuADS demonstrates performance nearly identical to that of CMA-ES in a three-dimensional setup. 
This structure makes optimization more straightforward for CMA-ES, leading to an increased $p$, which consequently reduces the influence of quadratic acceleration.
However, in scenarios with higher dimensions, where $p$ is likely to be lower, the benefits of quantum search would become more pronounced.

\begin{figure*}[thb]
\centering
\subfloat[Result for three dimensional rastrigin function.]{
\includegraphics[scale=0.35]{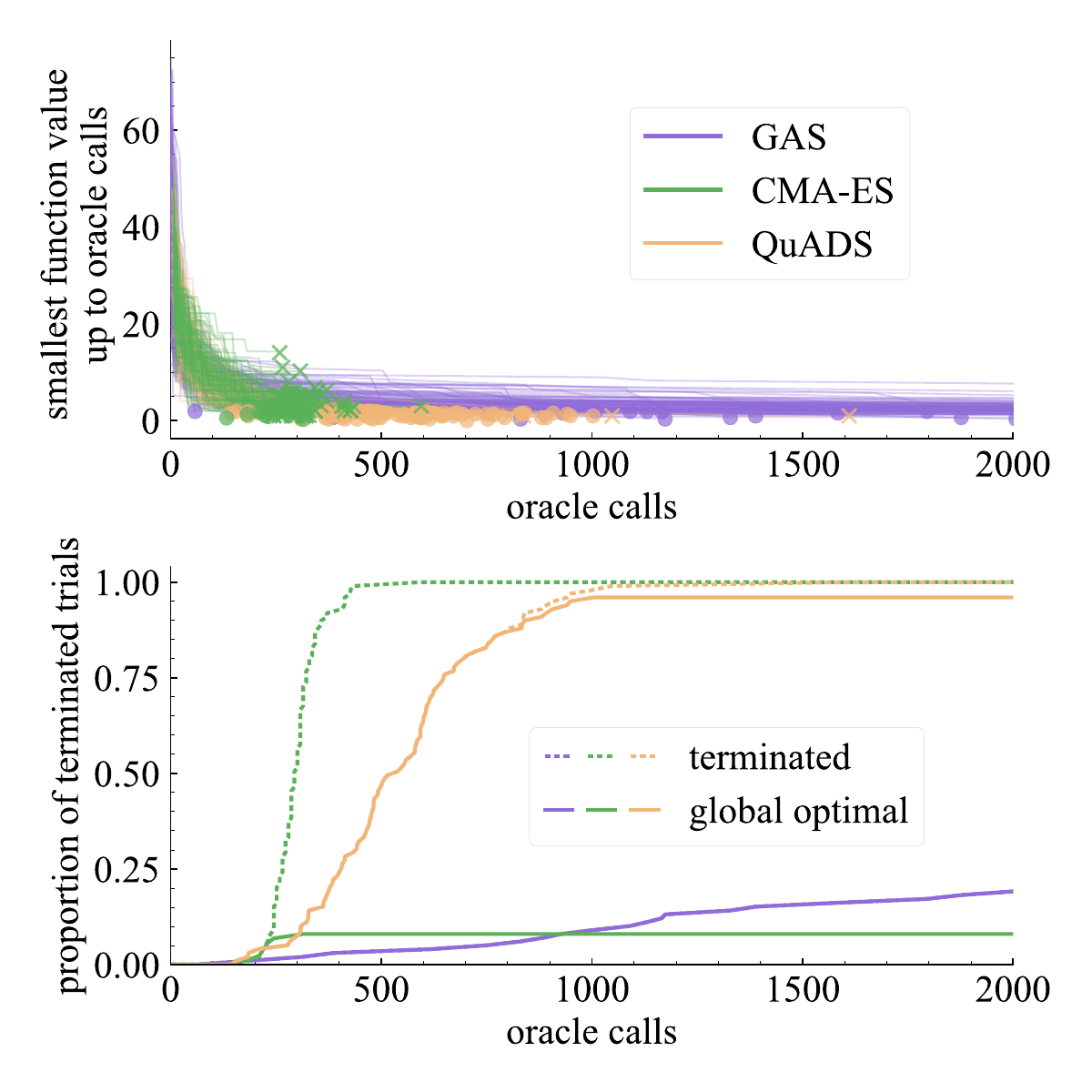}
\label{fig:quantum_rastrigin_3}
}
\hfil 
\subfloat[Result for three dimensional schwefel function.]{
\includegraphics[scale=0.35]{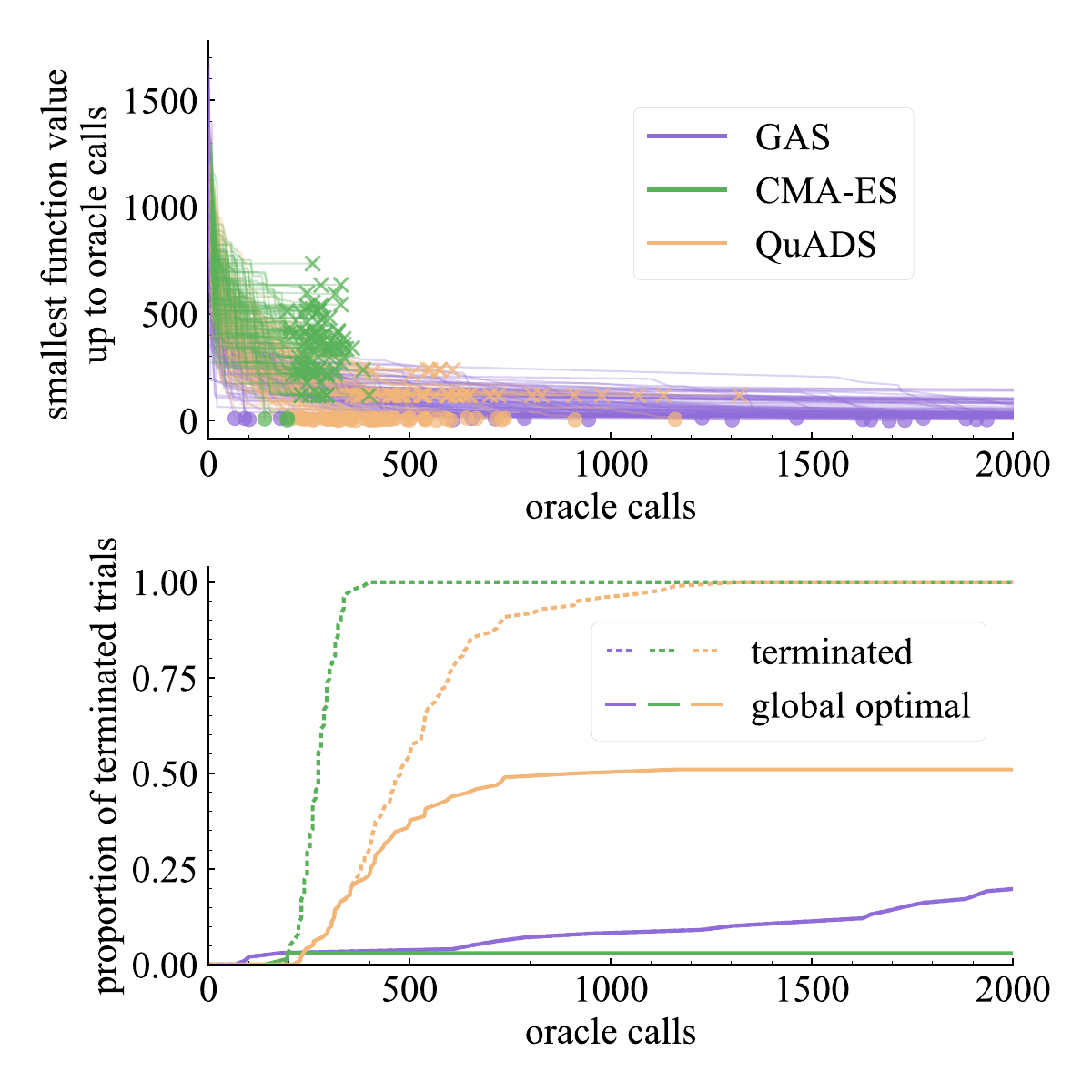}
\label{fig:quantum_schwefel_3}
}
\caption{Optimization processes for the three-dimensional rastrigin function (\ref{fig:quantum_rastrigin_3}) and schwefel function (\ref{fig:quantum_schwefel_3}). The upper figure presents the smallest function value obtained up to a specific number of oracle calls in each trial. A circle marker at the end of each trial represents the trial found a global solution, and a cross marker indicates local convergence. The lower figure shows the proportion of trials that found a global solution (solid line) and that reached the terminal condition (dashed line) up to the number of oracle calls.
The number of oracle calls required for local convergence can be larger than those for finding the global optimum.  
This arises from the algorithms' termination criteria: they terminate immediately upon locating the global optimum, but may continue until convergence of $\sigma$ if the global optimum remains unfound.
}
\label{fig:3d_optimization}
\end{figure*}

In Fig.~\ref{fig:quantum_rastrigin_3} and \ref{fig:quantum_schwefel_3}, we show the optimization process of the three-dimensional rastrigin function and schwefel function, for which the QuADS performs significantly better than the other methods.
Our results indicate that CMA-ES converges quickly but mostly fails to find a globally optimal solution; GAS finds a globally optimal solution on every trial, but its search is inefficient and requires a large number of oracle calls.
We observe that QuADS is more likely to converge to a global solution than CMA-ES and performs its search more efficiently than GAS.

\subsection{Performance estimation by classical algorithms}
\label{sec:exp_classic}
Next, we examine the performance of our method in higher dimensional settings.
As the dimensionality increases, the number of local optima grows exponentially, significantly complicating the optimization problem. 
Consequently, this leads to a reduction in $p$, enhancing the effectiveness of the quadratic acceleration inherent in quantum search. Therefore, it is anticipated that QuADS becomes increasingly advantageous under these conditions.
Since the size of the state vector in quantum simulators increases exponentially as the dimension increases, conducting simulations on a quantum simulator becomes difficult in more than 4 dimensions when $\tau = 8$.
To measure performance in higher dimensions, we replace the quantum sampling in QuADS and GAS with classical sampling and estimate the number of oracle calls required for the entire optimization in the case of quantum sampling from the optimization process of the corresponding classical algorithm.

For this, we utilize the optimal oracle call count, $N_{\mathrm{opt}}(p)$, of amplitude amplification,
\begin{align}
    N_{\mathrm{opt}}(p) = \frac{\arccos \sqrt p}{2 \arcsin \sqrt p},
    \label{eq:oracle_call_estimation}
\end{align}
where $p$ represents the probability of obtaining a correct sample in the classical sampling. 
When we set the number of rotations $r$ in amplitude amplification to $N_{\mathrm{opt}}(p)$, and for $\theta$ satisfying $\sin^2(\theta) = p$, we find that $\sin^2((2N_{\mathrm{opt}}(p) + 1)\theta) = 1$.
This result demonstrates that $N_{\mathrm{opt}}(p)$ indeed represents the optimal number of rotations.
Thus, $N_{\mathrm{opt}}(p)$ serves as a lower bound of the expected oracle call count within each amplitude amplification step of GAS or QuADS.
We estimate $p$ for each step by classical sampling; if obtaining $M_i$ samples under the threshold requires $K_i$ function evaluations at $i$th iteration of the corresponding classical algorithm, we estimate probability $p_i$ as $\tilde{p_i} = M_i / K_i$.
We then estimate a lower bound of $o_{\rm local}$ ($o_{\rm global}$) by an average of $\sum_{i} N_{\mathrm{opt}}(\tilde{p_i})$ using the optimization trials that achieved local (global) convergence.
Then, a lower bound to $o_{\rm total}$, which we denote as $\tilde o_{\rm lower}$, is estimated through Eq. \eqref{eq:expected_total_calls} using the corresponding estimated lower bounds for $o_{\rm local}$ and $o_{\rm global}$. 
Finally, $o_{\rm total}$ can be inferred from $\tilde o_{\rm lower}$, by multiplying a certain constant coefficient. This approach is valid because both $o_{\rm total}$ and $\tilde o_{\rm lower}$ scale in a similar manner with respect to $D$.
We conducted 100 individual optimizations to estimate $\tilde o_{\rm lower}$ for each of the rastrigin, schwefel, styblinski tang, ackley, alpine01, and griewank functions up to feasible dimensions in classical algorithms.
We choose the functions because QuADS surpasses GAS and CMA-ES greatly for the former two, while for the latter functions, differences between CMA-ES and QuADS are small (see Fig. \ref{fig:expected_global_eval}).

\begin{figure}[pthb]
\includegraphics[scale=0.4]{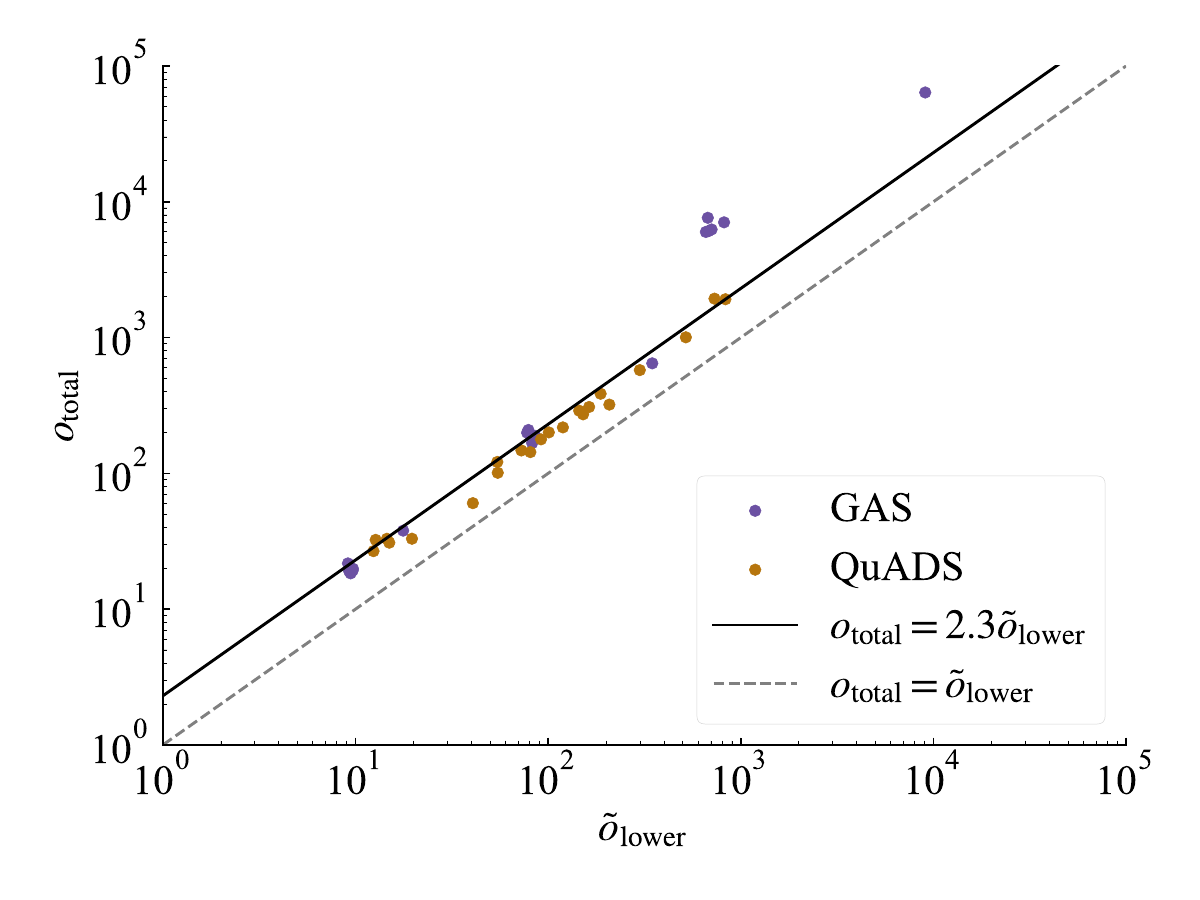} 
\caption{\label{fig:estimation_validity}
The consistency verification between classical estimation and quantum simulation regarding the expected number of oracle calls across the functions in Fig. \ref{fig:high-dim-exp}. 
The dotted line represents $o_{\rm total} = \tilde{o}_{\rm lower}$.
The classical estimation results ($\tilde o_{\rm lower}$) serve as a lower bound to the quantum simulation results ($o_{\rm total}$), which is consistent with theoretical predictions. 
The results of the quantum simulations are approximately twice that of the classical estimations, as indicated by the solid line $o_{\rm total}=2.3\tilde o_{\rm lower}$ for both QuADS and GAS. 
}
\end{figure}

First, we checked the consistency of using $\tilde o_{\rm lower}$ to estimate $o_{\rm total}$.
Figure \ref{fig:estimation_validity} shows the relationship between $\tilde{o}_{\rm{lower}}$ and $o_{\rm{total}}$.
Each data point in the plot corresponds to a specific function and dimension for which both $\tilde{o}_{\rm{lower}}$ and $o_{\rm{total}}$ can be computed.
It shows $o_{\rm total}$ is correctly lower bounded by $\tilde o_{\rm lower}$.
Furthermore, it can be observed from simulation that $o_{\rm total}$ can be closely approximated by $o_{\rm total} = 2.3\tilde o_{\rm lower}$.
Thus, we use $\tilde o_{\rm total}:= 2.3\tilde o_{\rm lower}$ as an estimator for $o_{\rm total}$ of GAS and QuADS in high dimensional setting. More detail analysis about consistency of this approximation in high dimension is shown in Appendix \ref{sec:consistency}.

\begin{figure*}[pbht]
\centering
\subfloat[rastrigin function.]{
\includegraphics[scale=0.4]{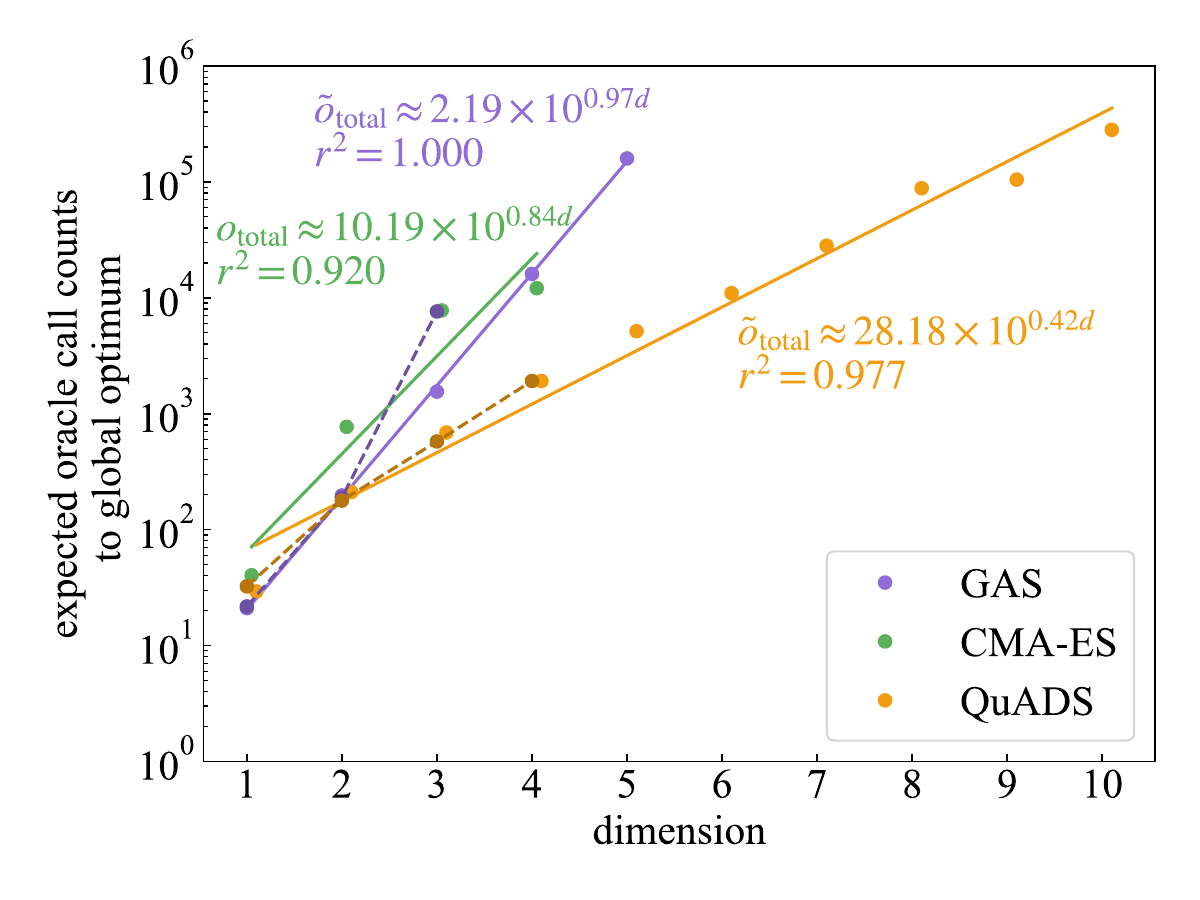}
\label{fig:high-dim-rastrigin}
}
\hfill 
\subfloat[schwefel function.]{
\includegraphics[scale=0.4]{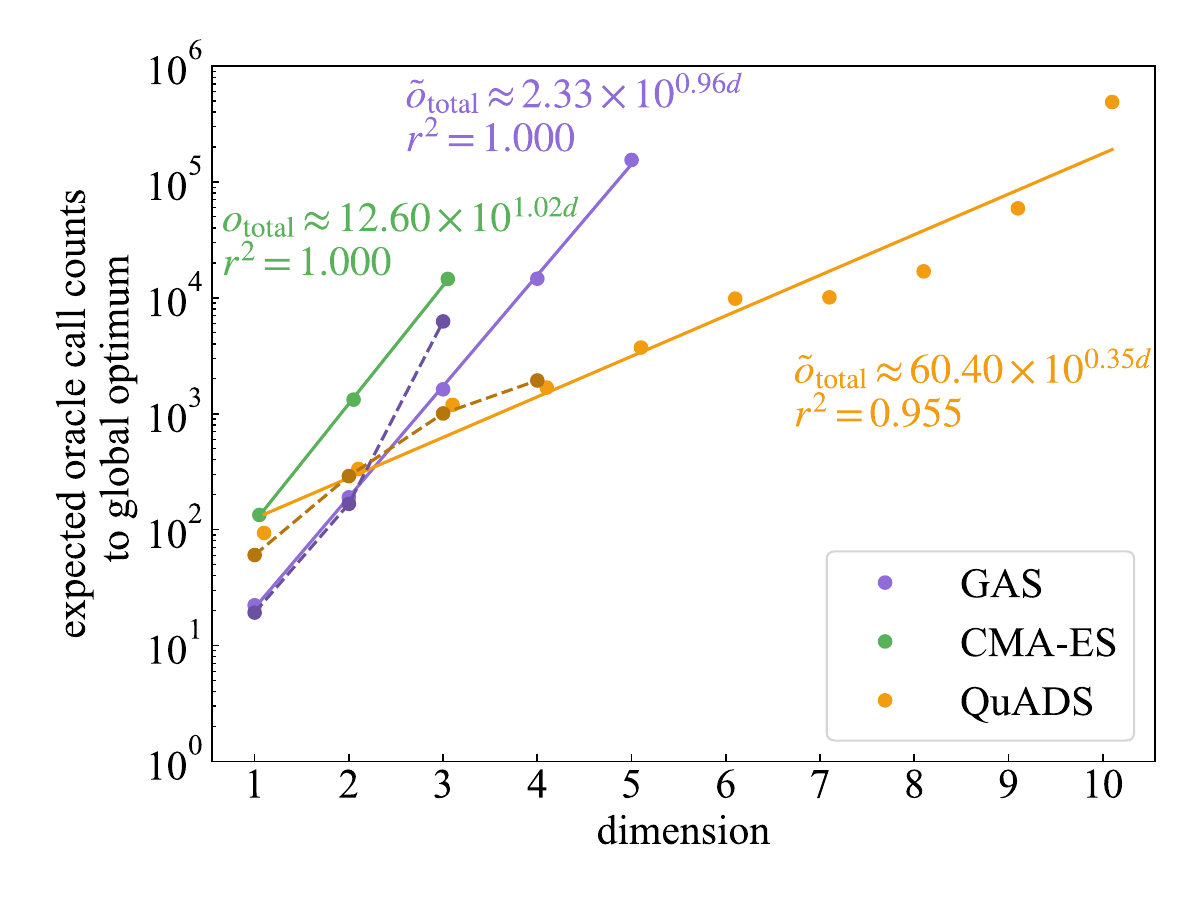}
\label{fig:high-dim-schwefel}
}
\\

\subfloat[styblinski tang function.]{
\includegraphics[scale=0.4]{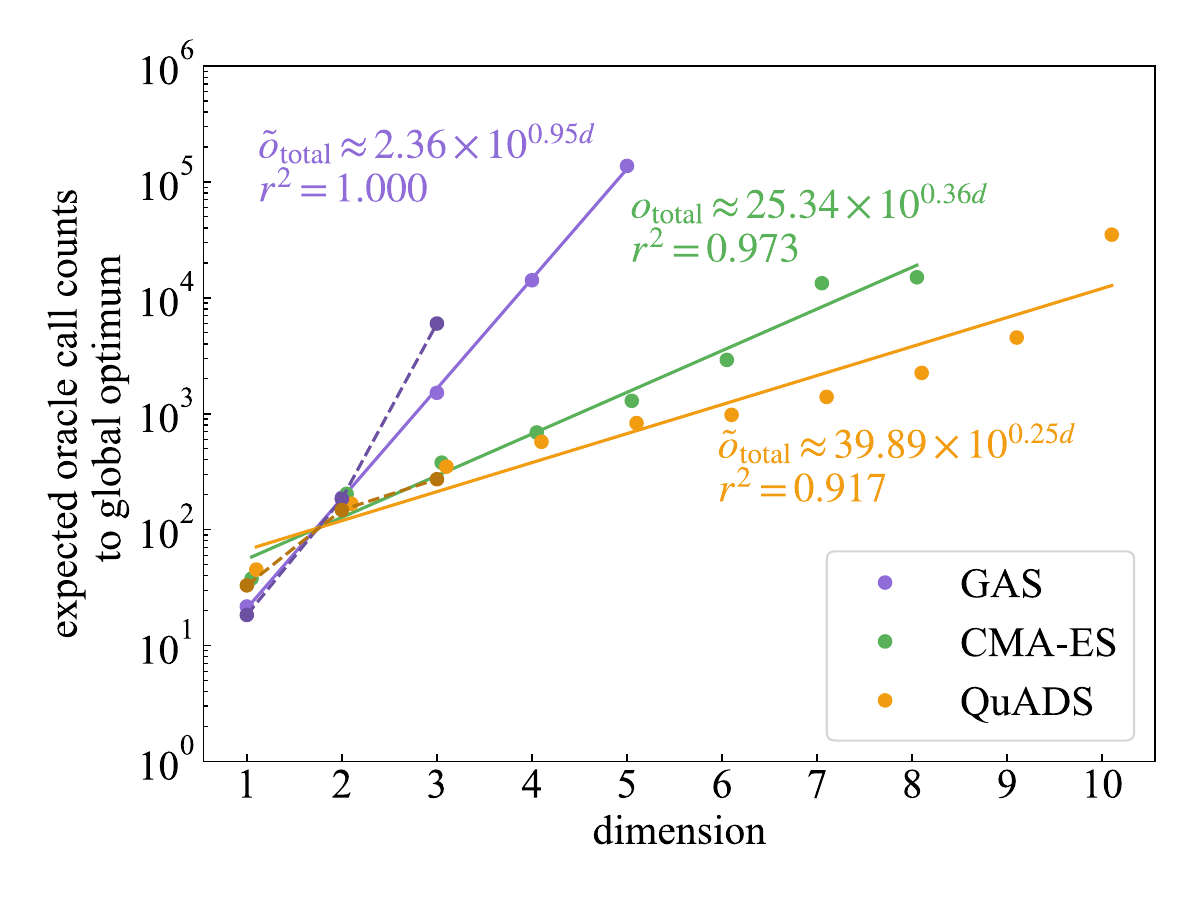}
\label{fig:high-dim-styblinski}
}
\hfill
\subfloat[ackley function.]{
\includegraphics[scale=0.4]{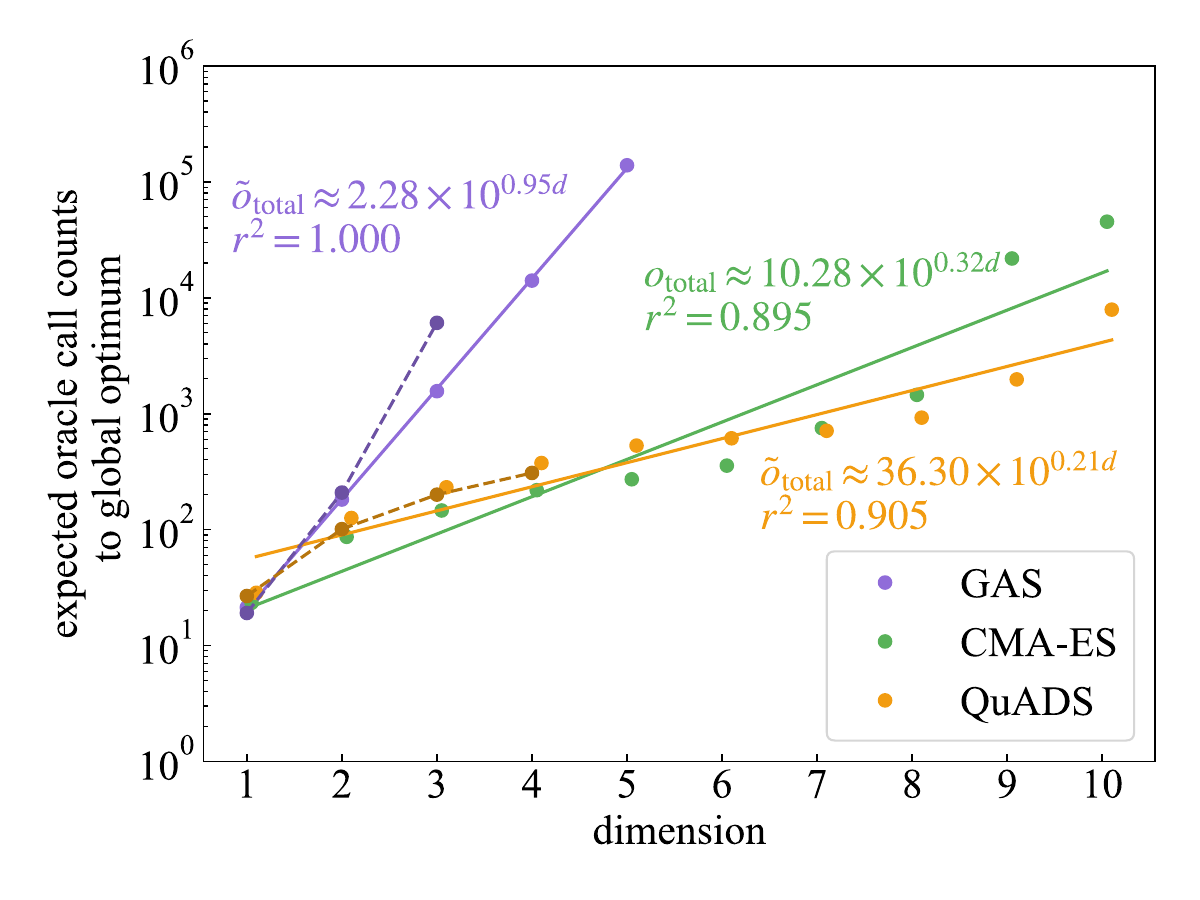}
\label{fig:high-dim-ackley}
}
\\

\subfloat[alpine01 function.]{
\includegraphics[scale=0.4]{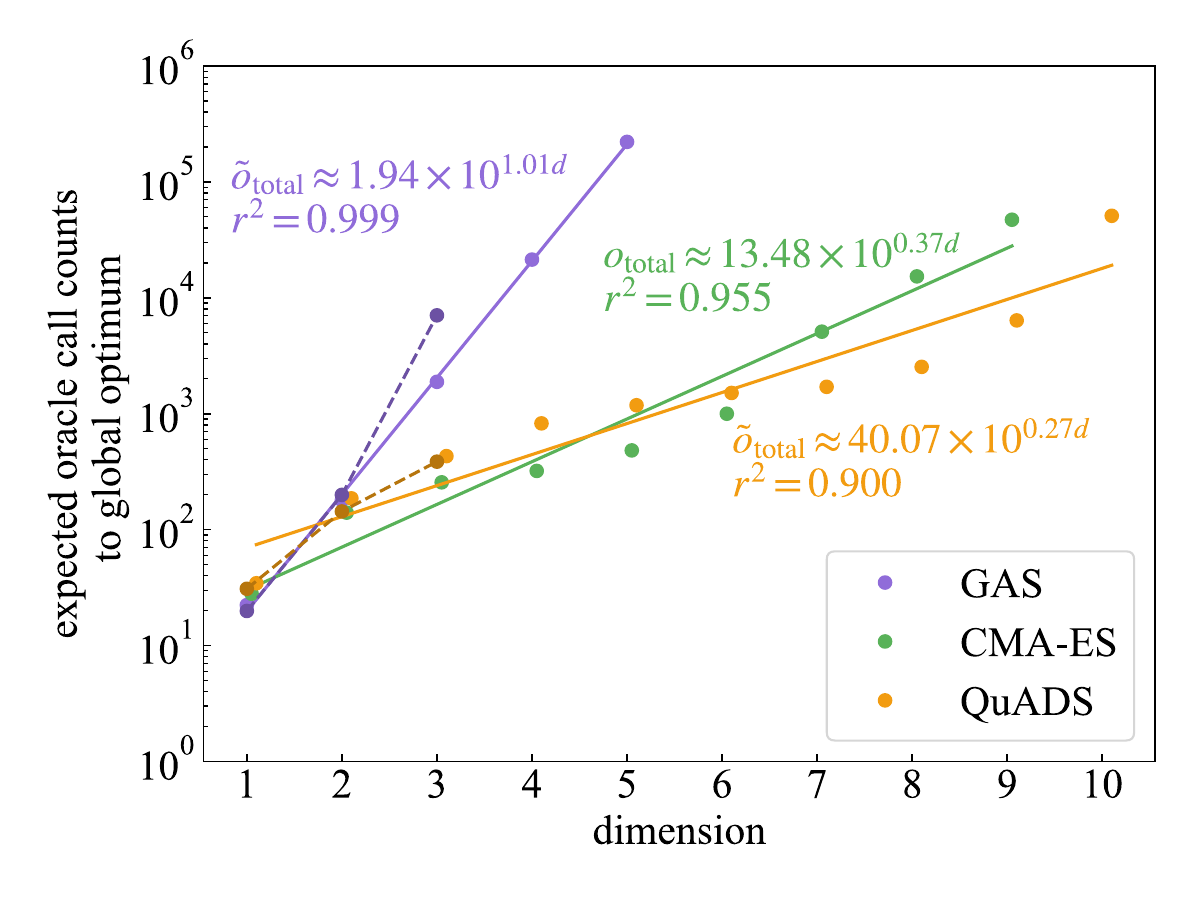}
\label{fig:high-dim-alpine01}
}
\hfill
\subfloat[griewank function.]{
\includegraphics[scale=0.4]{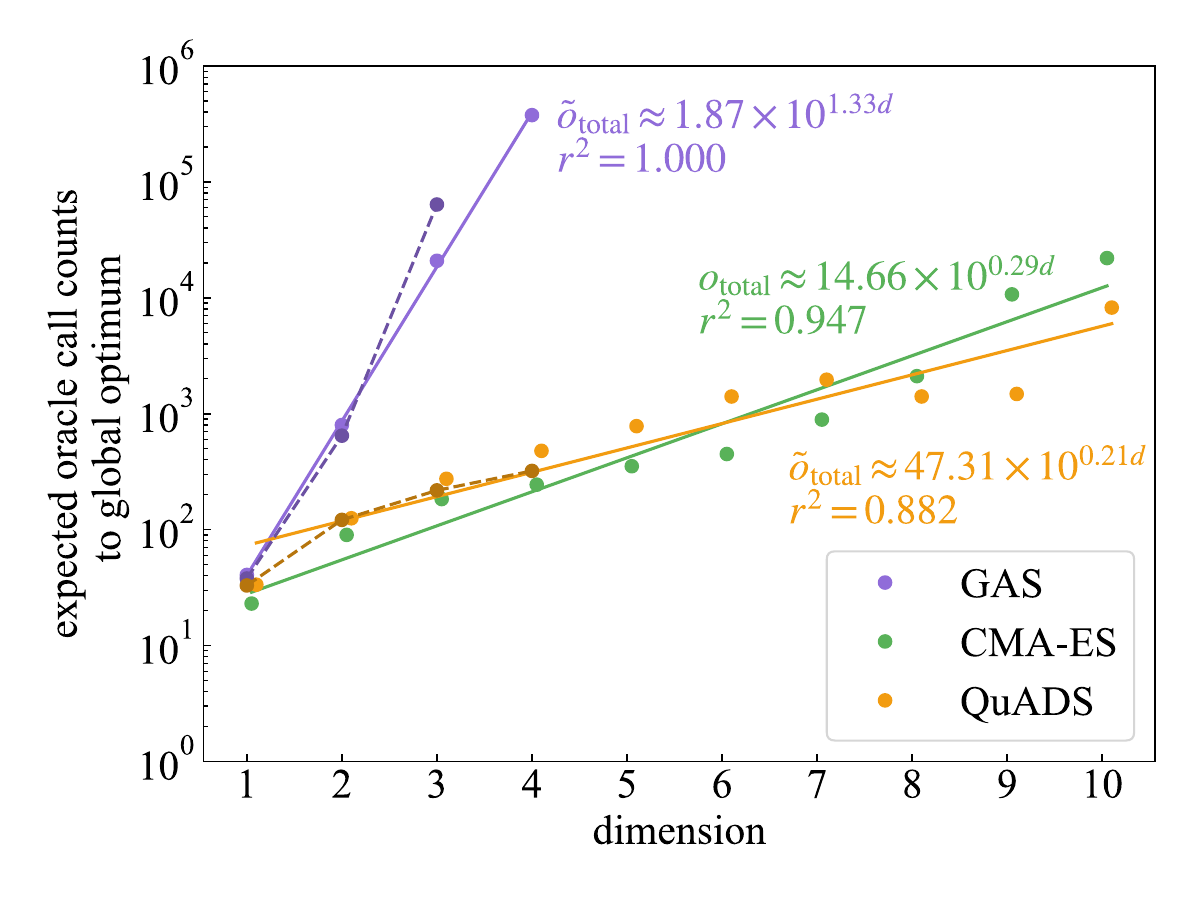}
\label{fig:high-dim-griewank}
}
\caption{\label{fig:high-dim-exp}Oracle call counts required to find the globally optimal solution estimated by running the corresponding classical optimization process. The solid line represents the regression line of $\tilde o_{\rm total}$ for GAS and QuADS and $o_{\rm total}$ for CMA-ES, and the dotted line represents $o_{\rm total}$ for GAS and QuADS. The equation of the regression line is displayed in the same color, and $r^2$ indicates the coefficient of determination for the regression. 
}
\end{figure*}

Figure~\ref{fig:high-dim-exp} presents the estimated $\tilde o_{\rm total}$ and $o_{\rm total}$ of GAS and QuADS and $o_{\rm total}$ of CMA-ES.
We found that QuADS consistently outperforms other methods in high-dimensional settings across all tested functions.
Furthermore, compared to other methods, QuADS exhibits a smaller increase in the number of oracle calls as the dimensionality increases.
Although GAS exhibits a nearly constant slope regardless of the function, CMA-ES and QuADS show distinct slopes for different functions. Notably, for the rastrigin and schwefel functions, QuADS achieves quadratic acceleration compared to CMA-ES.

\section{Conclusion}
\label{sec:conclusion}

This paper proposes a method for quantum continuous optimization in a quantum computer. In our method, we enhance GAS by integrating the efficient search capabilities of CMA-ES. 
Our method can efficiently utilize function structure in the search process through the updating rule of CMA-ES.
Our numerical simulation shows our method surpasses both CMA-ES and GAS in terms of expected oracle call count for finding global optimum. QuADS has a smaller increase in oracle call count with respect to dimension, and this observation suggests our method's significant potential in addressing high-dimensional problems commonly encountered in practical applications. 

We list possible future studies in the following.
First, although we neglected the cost for the initial state preparation, investigating the overall performance of QuADS including this cost is needed.
However, we expect that the state preparation cost can be negligible for complex objective functions $f(x)$.
Second, we can extend our approach to discrete variable cases.
Utilizing distribution in discrete optimization is generally referred to as ``estimation of distribution algorithm'' \cite{lozano_estimation_2002}.
Third, for the common challenge of continuous optimization on a quantum computer, we need to find practical oracles that are efficient in real time. The conditions for such an oracle would be multimodal functions which are difficult to optimize on a classical computer.
Also, oracles that are more efficient when computed on a quantum computer, like the energy of materials, are attractive candidates.


\section{Acknowledgement}
K. Morimoto and YT thanks IPA for its support through MITOU
Target program.
K. Mitarai is supported by JST PRESTO Grant No. PMJPR2019.
This work is supported by MEXT Quantum Leap Flagship Program (MEXTQLEAP) Grant No. JPMXS0118067394 and JPMXS0120319794. We also acknowledge support from JST COI-NEXT program Grant No. JPMJPF2014.


\appendix
\section{Hyperparameters for CMA-ES Update}
\label{sec:hyper}
\newcommand{\nsample}{K}

\begin{table}[h]
\centering
\begin{tabular}{c|c}

Variable & \\ \hline
$ w_i $ & $ \frac{\log(\nsample+1/2) - \log(i)}{\sum_{i=1}^K \nlr{\log(\nsample+1/2) - \log(i)}} $ \\
$ \mu_{\text{eff}} $ & $ \frac{1}{\sum_{i=1}^K w_i^2} $ \\
$ c_1 $ & $ \frac{2}{(D+1.3)^2 + \mu_{\text{eff}}} $ \\
$ c_\sigma $ & $ \frac{\mu_{\text{eff}} + 2}{D + \mu_{\text{eff}} + 5} $ \\
$ c_\mu $ & $ \min\left(1-c_1, 2 \left(\mu_{\text{eff}} - 2 + \frac{1}{\mu_{\text{eff}}}\right)\right) $ \\
$ d_\sigma $ & $\min\left(1-c_1, \frac{2(\mu_{\text{eff}} - 2 + \frac{1}{\mu_{\text{eff}}})}{(D + 2)^2 + \mu_{\text{eff}}}\right)$ \\
\end{tabular}
\caption{\label{tab:hyper}Hyperparameters setting for CMA-ES update}
\end{table}

In this section, we describe hyperparameters used in Section \ref{sec:experiments}.
For CMA-ES update (Algorithm \ref{alg:cma-update}) in CMA-ES and QuADS, we employ the default hyperparameters used in \cite{hansen_cma_2023}. 
The hyperparameters for CMA-Update are enumerated in Table \ref{tab:hyper}.

\section{Particle swarm optimization and basin hopping optimizer}
\label{sec:PSO}
PSO \cite{Kennedy_particle, Shi_modified} is a swarm-based optimizer that iteratively improves candidate solutions through communication among the swarm. PSO simulates the social behavior of birds within a flock. The position of each bird in the search space represents a potential solution, and its flight is guided by its own experience as well as the experience of other birds.
The update equations for the velocity and position of each particle can be written as:
\begin{align*}
    v_{i}^{(t+1)} &= w v_{i}^{(t)} + c_1 r_1 (p_{i} - x_{i}^{(t)}) + c_2 r_2  (p_{g} - x_{i}^{(t)}) \\
    x_{i}^{(t+1)} &= x_{i}^{(t)} + v_{i}^{(t+1)}
\end{align*}
where $v_{i}^{(t)}$ is the velocity of particle $i$ at iteration $t$, $x_{i}^{(t)}$ is the current position of particle $i$, $p_{i}$ is the best known position of particle $i$, $p_{g}$ is the global best known position among all particles, $w$ is the inertia weight, $c_{1}$ and $c_{2}$ are the personal and global coefficients, respectively, and $r_{1}$ and $r_{2}$ are random numbers between 0 and 1.
For the implementation of PSO, the PySwarm library \cite{pyswarmsJOSS2018} was utilized, with the hyperparameters set as $w=0.9$, $c_1=0.5$, and $c_2=0.3$.

Basin-hopping is a global optimization algorithm that enhances a local optimization method through the integration of stochastic perturbations to the coordinates, enabling the algorithm to escape local minima. In our implementation, the basinhopping module from SciPy optimization library\cite{2020SciPy-NMeth} was employed. Due to the default configuration does not handle bounds on the optimization function, we modified the random perturbation process to ensure it does not exceed the defined bounds. If a perturbation results in exceeding the bounds, the solution is adjusted back to the nearest boundary. As a subroutine of the basinhopping algorithm, the L-BFGS-B method \cite{byrd_limited_1995} was adopted.

\section{Test functions}
\label{sec:test-func}
\newcommand{\funcsimagescale}{0.33}

\begin{figure*}
\subfloat[rastrigin function.]{
\includegraphics[scale=\funcsimagescale]{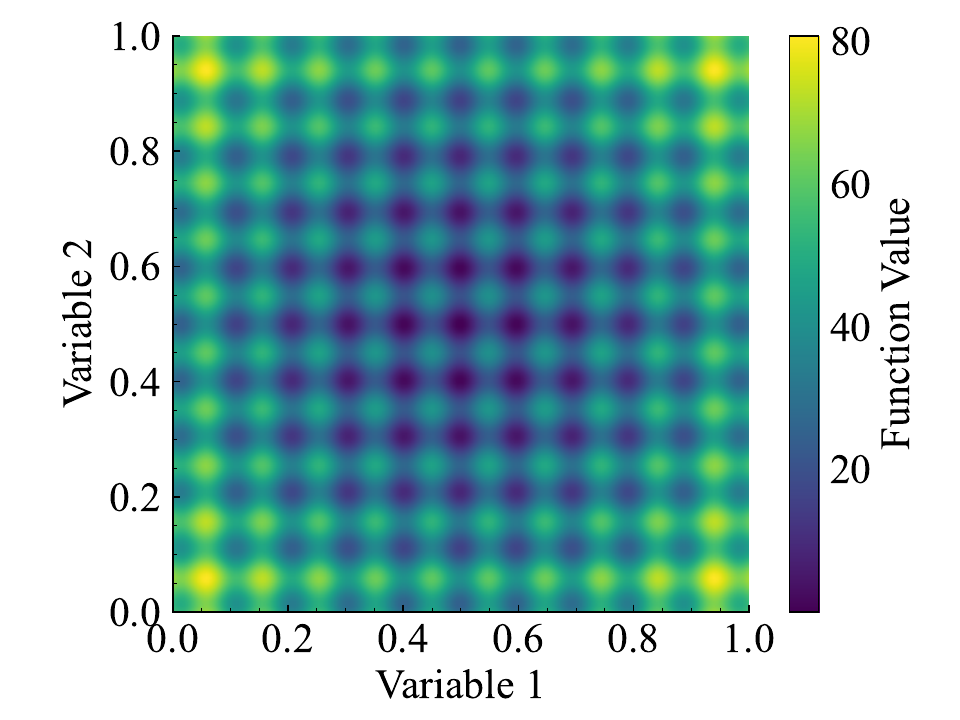}
\label{fig:func-rastrigin}
}
\hfill
\centering
\subfloat[ackley function.]{
\includegraphics[scale=\funcsimagescale]{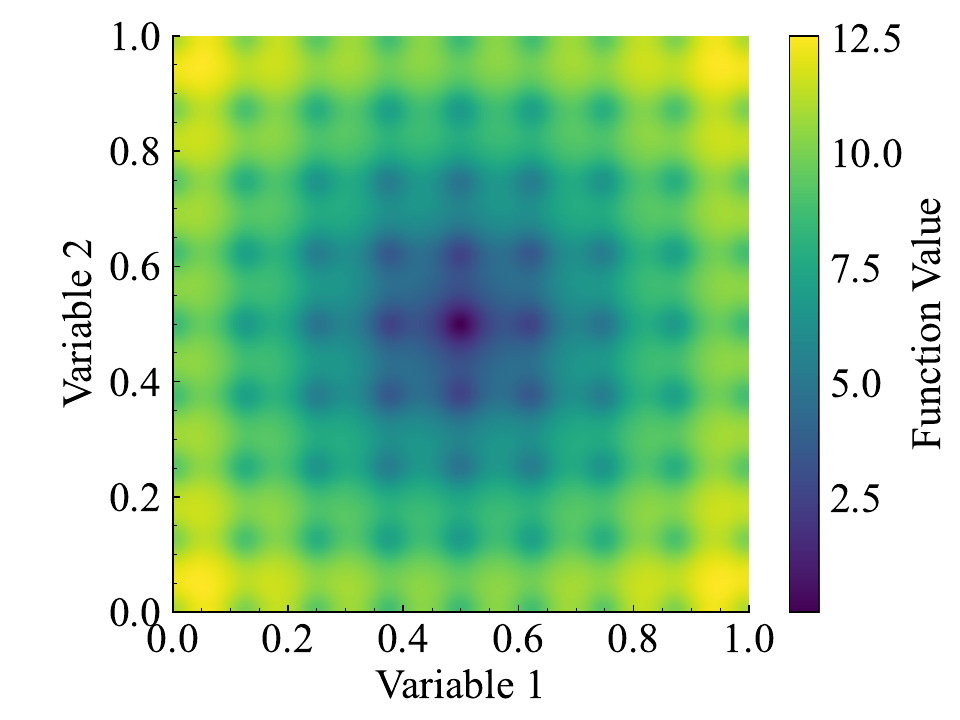}
\label{fig:func-ackley}
}
\hfill 
\subfloat[alpine01 function.]{
\includegraphics[scale=\funcsimagescale]{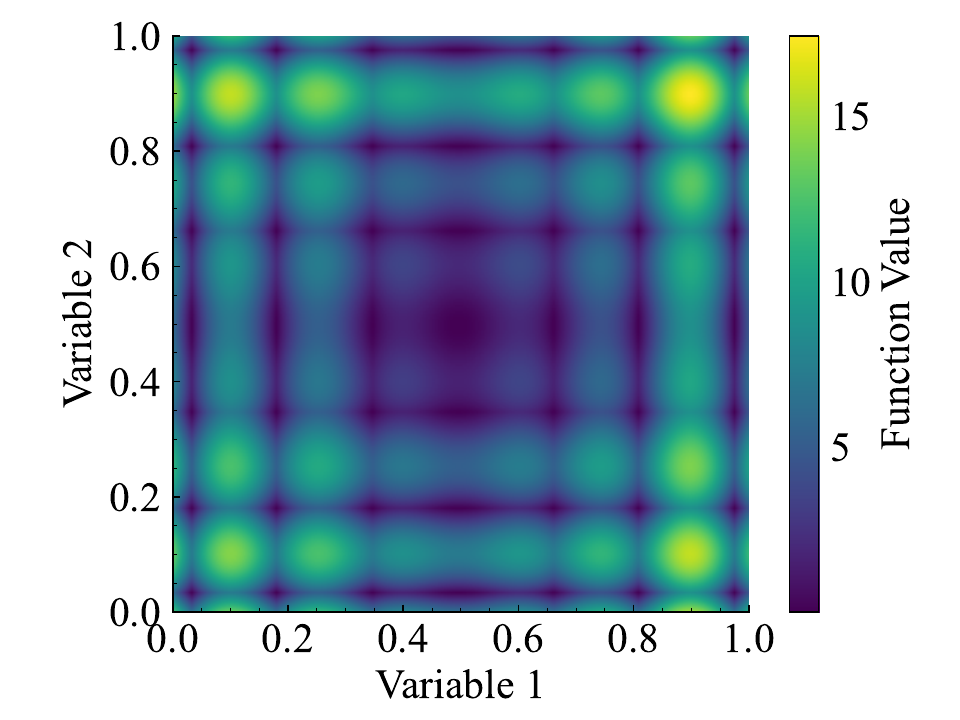}
\label{fig:func-alpine01}
}
\\
\subfloat[alpine02 function.]{
\includegraphics[scale=\funcsimagescale]{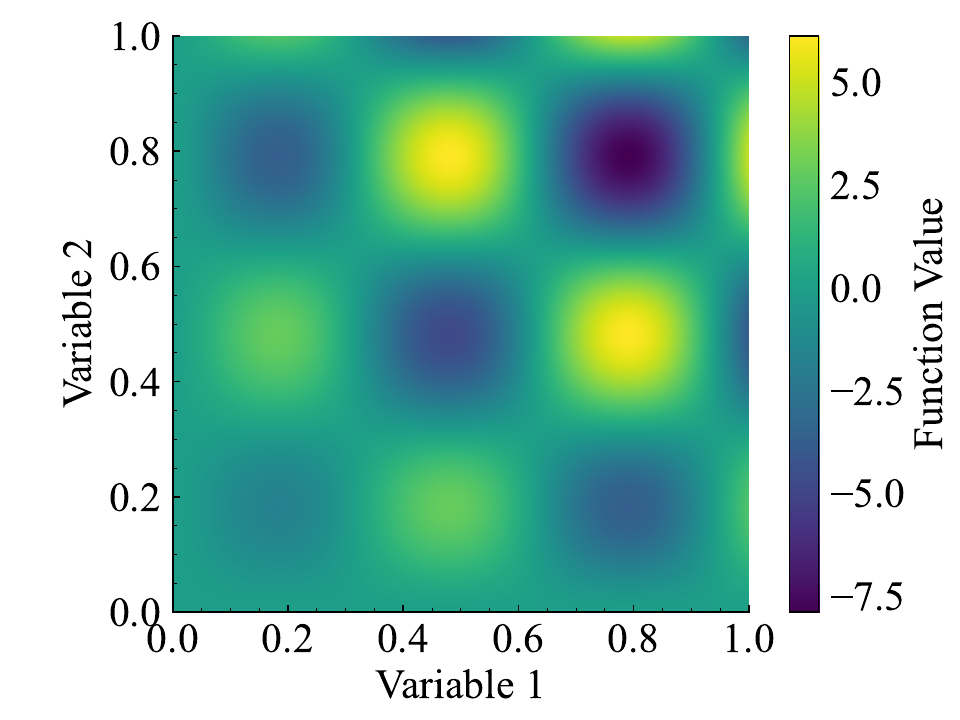}
\label{fig:func-alpine02}
}
\hfill
\subfloat[deflected corrugated spring function.]{
\includegraphics[scale=\funcsimagescale]{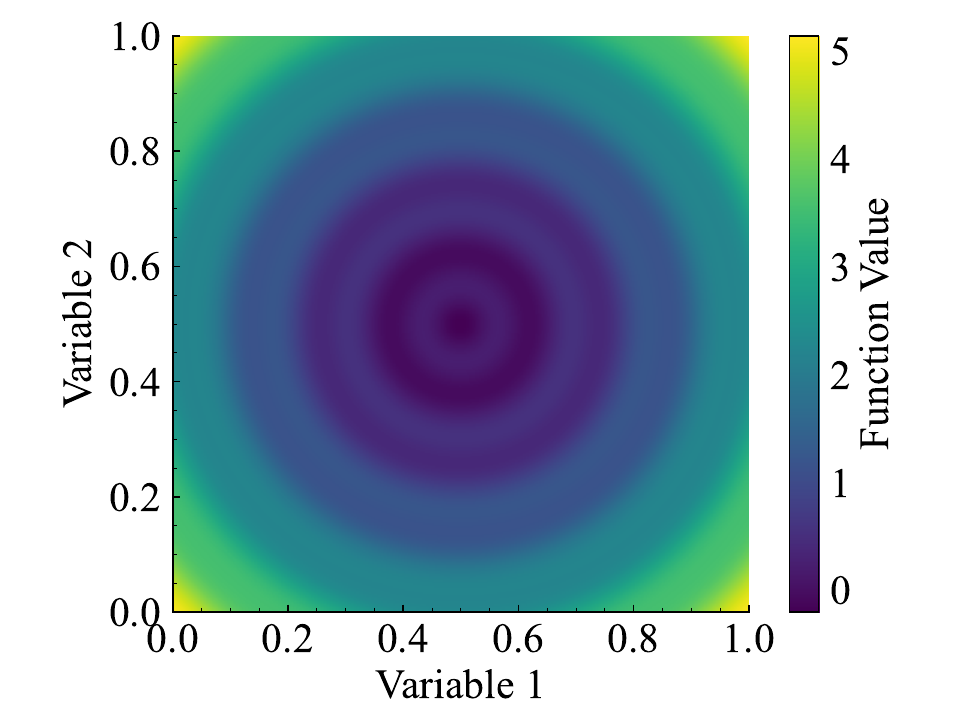}
\label{fig:func-deflectedCorrugatedSpring}
}
\hfill
\subfloat[griewank function.]{
\includegraphics[scale=\funcsimagescale]{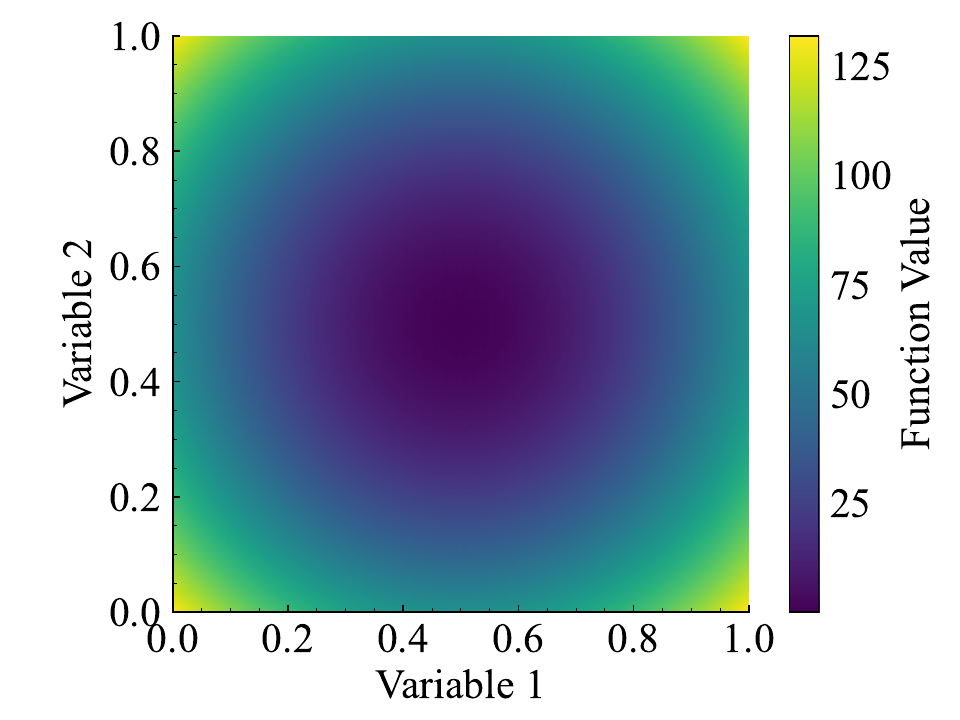}
\label{fig:func-griewank}
}
\\
\subfloat[schwefel function.]{
\includegraphics[scale=\funcsimagescale]{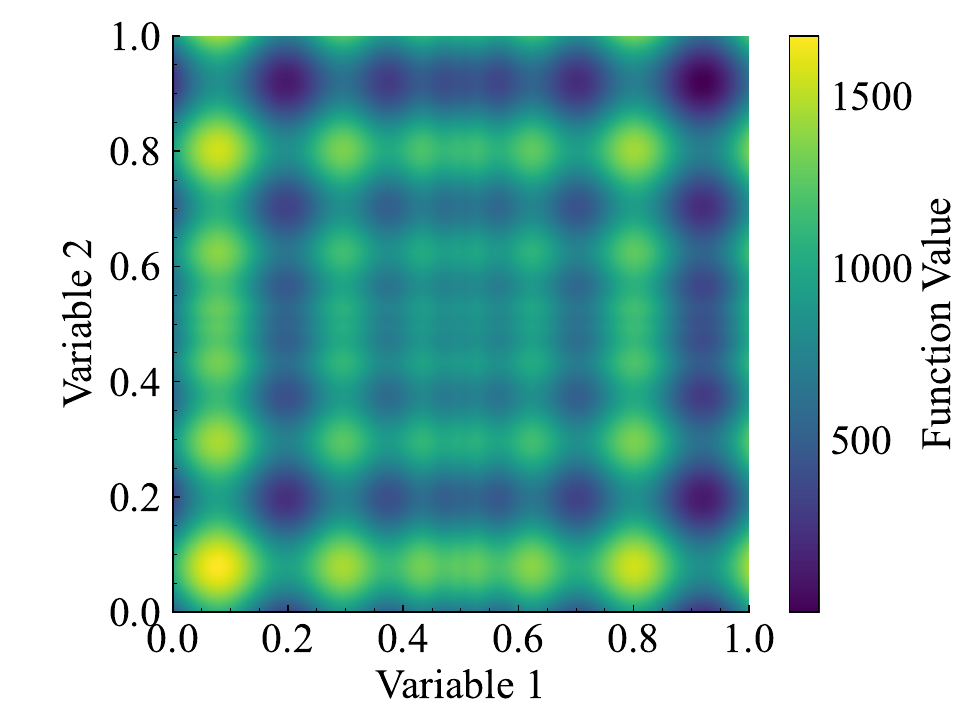}
\label{fig:func-schwefel}
}
\hfill
\subfloat[styblinski function.]{
\includegraphics[scale=\funcsimagescale]{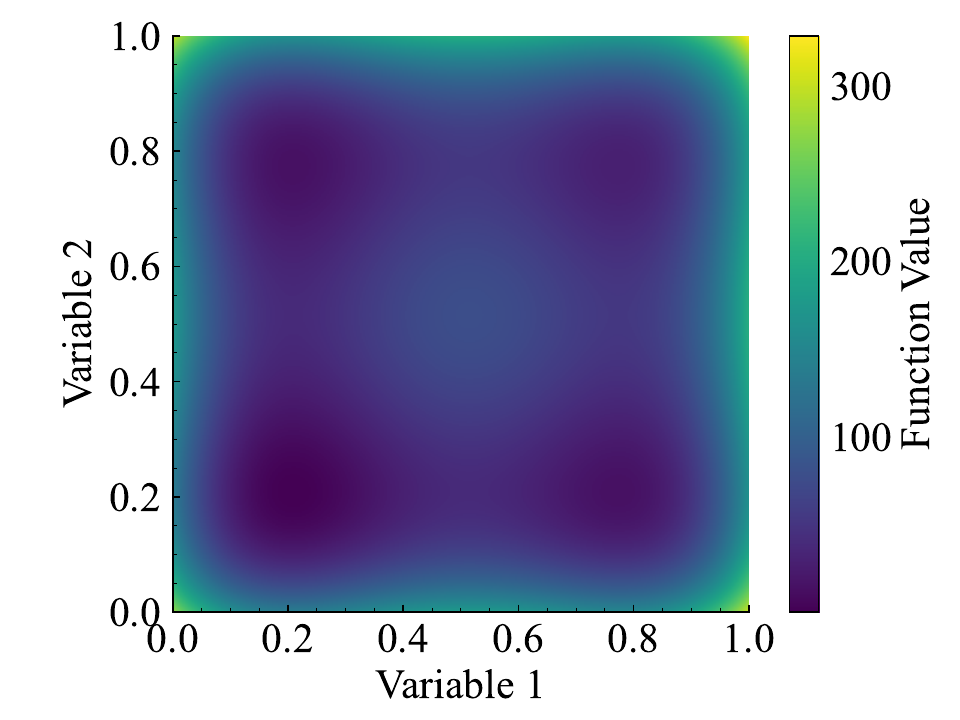}
\label{fig:func-styblinski}
}
\hfill
\subfloat[wavy function.]{
\includegraphics[scale=\funcsimagescale]{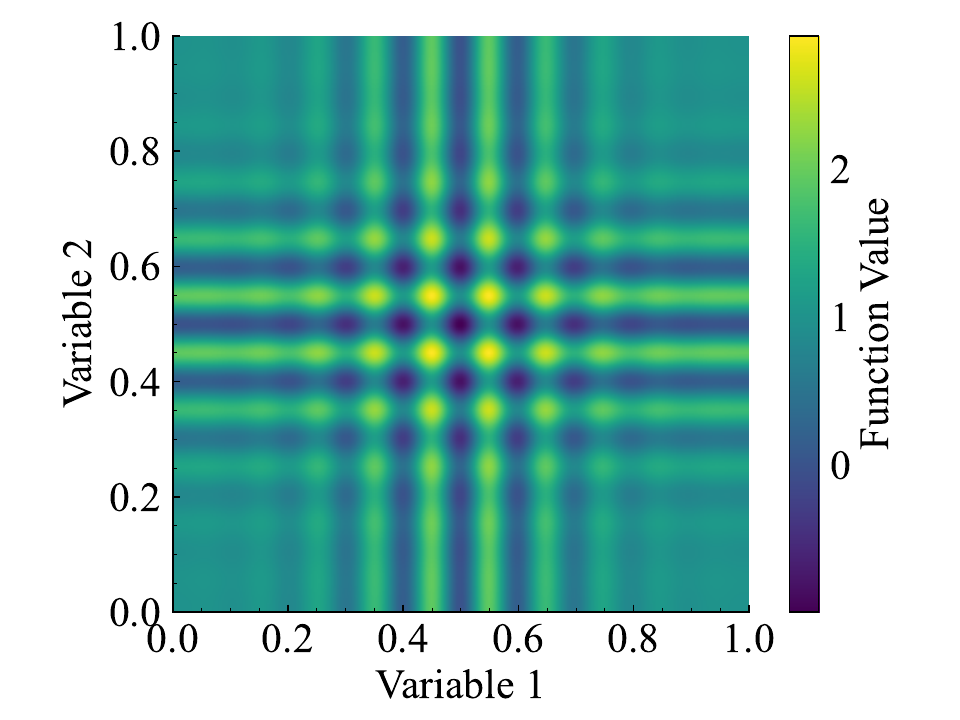}
\label{fig:func-wavy}
}
\hfill
\caption{\label{fig:funcs_figure}The landscapes of two dimensional test functions in Table \ref{tab:test-func}.}
\end{figure*}

\begin{figure}
    \centering
\includegraphics[scale=0.4]{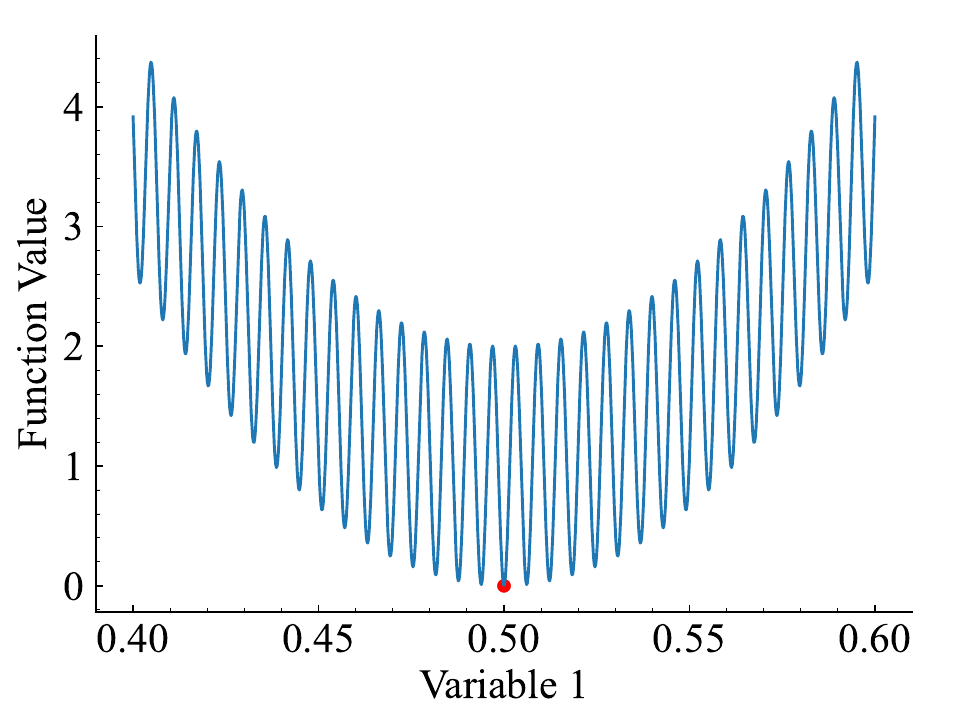}
    \caption{The landscape of one dimensional griewank function. The red dots indicate the minimum point of the griewank function.}
    \label{fig:griewank_1d}
\end{figure}

The plots of test functions in Table \ref{tab:test-func} are shown in Figure \ref{fig:funcs_figure} for the two-dimensional case.
Although the griewank function looks similar to a quadratic function in Figure \ref{fig:funcs_figure}, it actually exhibits significant fluctuations. 
A more detailed view of griewank function at a smaller scale is provided in Figure \ref{fig:griewank_1d}.

\section{Consistency of approximation of the number of oracle calls with classical simulations}
\label{sec:consistency}
In Section \ref{sec:exp_classic}, we estimate oracle counts of quantum computation in high dimension from the number of function calls obtained from classical computations using the equation $o_{\rm total} = 2.3 \tilde o_{\rm lower}$. Does the relation hold true for higher dimensions as well? This section aims to demonstrate the validity of this transformation in higher dimension.
Specifically, we investigated the ratio of the expected number of oracle calls required for amplitude amplification to $N_{opt}(p)$ when $\lambda=5/4$ for each $p$. 

\begin{figure}
\includegraphics[scale=0.4]{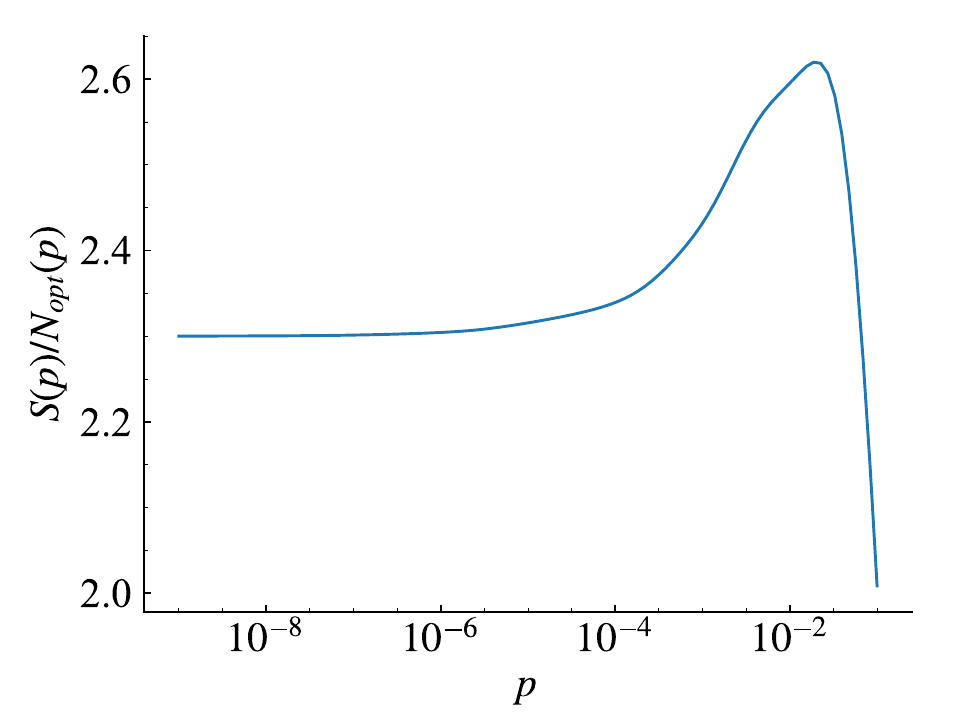} 
\caption{\label{fig:consistency_of_coeff}
$S(p) / N_{opt}(p)$ for each $p$.
}
\end{figure}

First, let's derive the expression for the expected number of oracle calls in amplitude amplification when $p$ is fixed.
The maximum rotation count $n_k$ in the $k$-th trial is given by
$n_k := \left\lfloor \lambda^{k-1} \right\rfloor$ and the acceptance probability $a_r$ after performing $r$ amplitude amplifications is
\begin{align*}
a_r := \sin^2\left( \left( 2 r + 1\right) \arcsin{\sqrt{p}} \right).
\end{align*}
In the $k$-th trial, we uniformly select $r$ from integers $0,\ldots,\left\lfloor \lambda^{k-1} \right\rfloor$ and perform $r$ amplitude amplifications. 
Thus, the probability of rejection $b_k$ in the $k$-th trial is expressed as
\begin{align*}
b_k := 1 - \frac{1}{n_k + 1}\sum_{r = 0}^{n_k} a_r.
\end{align*}
Therefore, the expected total number of oracle calls $S$ required until acceptance is computed as follows:
\begin{align*}
S(p) &= \lim_{K \to \infty} \sum_{k = 1}^{K} \sum_{r = 0}^{n_k} \frac{1}{n_k + 1} r\left(\prod_{j=1}^{k-1} b_j\right)a_r\\ &= \lim_{K \to \infty} \sum_{k = 1}^{K} \frac{1}{n_k + 1} \left(\prod_{j=1}^{k-1} b_j\right) \left(\sum_{r = 0}^{n_k} r a_r\right).
\end{align*}
We compute this sum of infinite series by finite approximation with a sufficiently large $K$. 
Figure \ref{fig:consistency_of_coeff} shows the ratio $S(p) / N_{opt}(p)$ for each $p$. 
This ratio converges around $2.3$ when $p$ is small, which is consistent with Fig. \ref{fig:estimation_validity}.

\bibliography{references}

\end{document}